\documentclass[aps,prb,twocolumn,nofootinbib,longbibliography,superscriptaddress]{revtex4-2}
\usepackage{graphicx}
\usepackage{amsfonts}
\usepackage{amscd}
\usepackage{amsmath}
\usepackage{amssymb}
\usepackage{newtxtext,newtxmath}
\usepackage{here}
\usepackage{tabularx}
\usepackage{etoolbox}
\usepackage{bm}
\usepackage{mathrsfs}
\usepackage{listings}
\usepackage{color}
\usepackage{braket}
\usepackage[normalem]{ulem}  
\usepackage{hyperref}

\newcommand{\nn}{\nonumber}
\newcommand{\beq}{\begin{eqnarray}}
\newcommand{\eeq}{\end{eqnarray}}

\newcommand\erase{\bgroup\markoverwith{\textcolor{blue}{\rule[.5ex]{2pt}{1pt}}}\ULon}

\begin{document}
\title{Intrinsic spin Nernst effect in spin-triplet superconductors}

\author{Taiki Matsushita$^\ast$}
\affiliation{Department of Materials Engineering Science, The University of Osaka, Toyonaka, Osaka 560-8531, Japan}
\affiliation{Spintronics Research Network Division, Institute for Open and Transdisciplinary Research
Initiatives, The University of Osaka, Japan}
\author{Youichi Yanase}
\affiliation{Department of Physics, Graduate School of Science, Kyoto University, Kyoto 606-8502, Japan}
\author{Takeshi Mizushima}
\affiliation{Department of Materials Engineering Science, The University of Osaka, Toyonaka, Osaka 560-8531, Japan}
\author{Satoshi Fujimoto}
\affiliation{Department of Materials Engineering Science, The University of Osaka, Toyonaka, Osaka 560-8531, Japan}
\affiliation{Spintronics Research Network Division, Institute for Open and Transdisciplinary Research
Initiatives, The University of Osaka, Japan}
\affiliation{Center for Quantum Information and Quantum Biology, The University of Osaka, Toyonaka, Osaka 560-8531, Japan}
\author{Ilya Vekhter}
\affiliation{Department of Physics and Astronomy, Louisiana State University, Baton Rouge, LA 70803-4001, USA}
\date{\today}


\begin{abstract}
We theoretically investigate the intrinsic (impurity-independent) spin Nernst effect (SNE), a spin current generation perpendicular to temperature gradients, in spin-triplet superconductors.
We show that, in these systems, the SNE consists of two distinct contributions: a direct quasiparticle contribution and an indirect supercurrent contribution.
The quasiparticle contribution originates from the momentum space Berry curvature generated by spin-triplet Cooper pairs.
The indirect contribution arises from a compensating supercurrent that cancels the bulk thermoelectric charge current.
While this contribution vanishes when the condensate has no spin-polarization in momentum space, it can be comparable in magnitude to the quasiparticle contribution in nonunitary superconductors.
These results demonstrate that thermoelectric spin supercurrent must be explicitly accounted for when evaluating the SNE in nonunitary superconductors.
\end{abstract}
\maketitle

\section{Introduction}
\label{sec: intr}
Since the discovery of the quantum Hall effect, intrinsic (impurity-independent) transverse transport phenomena have been widely recognized as signatures of topological electronic states~\cite{hasan_TI,qi_TI,sato2016majorana}. 
In superconductors, where the charge response is shunted by the supercurrent, the corresponding signatures appear in thermal responses. 
The anomalous thermal Hall effect (ATHE), a thermal current generation normal to temperature gradients in the absence of external magnetic fields, is regarded as a probe of topological superconducting phases with broken time-reversal symmetry~\cite{read_FQHE,sumiyoshi_ATHE,goswami_ATHE,goswami_ATHE2,moriya_ATHE,meng_ATHE}.

More recently, it has been shown that spin-current responses driven by temperature gradients provide complementary probes of topological superconductivity.
The extrinsic (impurity-induced) spin Nernst effect (SNE), a spin current generation normal to temperature gradients, has been proposed as a sensitive probe of the helical (spin-dependent chiral) superconducting order parameter, which realizes time-reversal-invariant topological superconductivity~\cite{matsuhita_SNE1}.
Although its intrinsic mechanism has been analyzed for non-topological (Rashba) superconductors, the analysis of the intrinsic SNE and its competition with extrinsic SNE is still lacking for topological superconductors~\cite{liao_conSC}.

In transverse current responses such as ATHE and SNE, the extrinsic (impurity-induced) and intrinsic contributions coexist and often compete~\cite{nagaosa_AHE,sinitsyn_AHE,fujimoto_SNE}.
This competition can obscure direct signatures of nontrivial band topology, making it essential to disentangle these.
Such competition has been extensively studied for ATHE in superconductors~\cite{yip2016low,yilmaz2020spontaneous,ngampruetikorn2020impurity,sharma_ATHE,wave24,matsushita_ATHE}, but the corresponding analysis for the SNE is absent.

In this work, we fill this gap by analyzing the intrinsic SNE in spin-triplet superconductors. This effect arises from the Berry curvature generated by spin-triplet Cooper pairing~\cite{xiao_berrycur}. 
We evaluate the intrinsic contribution to the spin Nernst conductivity (SNC) for helical and nonunitary superconducting states, and show that, although it is smaller than the corresponding extrinsic contribution for moderately clean superconductors, its magnitude lies within the measurable range for clean systems.

Notably, we demonstrate that, when evaluating the SNC in spin-triplet superconductors, it is essential to use the definition of the conserved spin current, which satisfies the continuity equation and includes the spin torque dipole~\cite{shi_conSC,zhang_conSC,murakami_conSC,xiao_mott}. 
We explicitly show that, in many situations where the conventional definition of the spin current predicts the absence of the SNE, employing the conserved spin current operator yields a finite SNC.

We also emphasize that the intrinsic SNE consists of two distinct contributions: a quasiparticle contribution and a complementary contribution arising from the supercurrent~\cite{matsushita_SSSE}.
Quasiparticles, which carry both entropy and spin, naturally couple to temperature gradients and thus generate the former~\cite{liao_conSC}.
Because Cooper pairs do not carry entropy, it has long been widely--though incorrectly--believed that the supercurrent does not contribute to thermal responses such as the SNE.
However, quasiparticles also carry electric charge, and their coupling to temperature gradients induces a thermoelectric charge current~\cite{shelly_resolving_2016}.
In the Meissner state, when a superconductor expels the magnetic flux from the bulk ($\bm{B}=0$), the total bulk charge current must vanish~\cite{ginzburg1978thermoelectric}. 
This necessitates a compensating thermoelectric supercurrent that cancels the quasiparticle thermoelectric charge current.
Crucially, if the condensate is partially spin-polarized, cancellation of the charge current does not eliminate the spin current~\cite{matsushita_SSSE}.
A net thermoelectric spin current therefore remains.
In a related development, Ref.~\cite{hsieh2025intrinsic} recently proposed detecting intrinsic quasiparticle Nernst currents in superconductors using a ring-shaped geometry, through the magnetic flux generated by the resulting circulating current.

In this paper, we focus on spin-triplet states as prototypical topological superconductors because they support nontrivial band topology over a wide range of parameters according to the Sato--Fu--Berg theorem~\cite{sato_TSC,sato_TSC1,fu_TSC}.
A number of candidate spin-triplet superconductors have been proposed so far, but identifying their superconducting order parameters remains a challenging problem~\cite{robert_UPt3,nomoto_UPt3,machida_UPt3,maeno_Sr2RuO4,aoki_FMSC,aoki_UTe2,machida_UBe13,sigrist_UBe13,shimizu_UBe13,yonezawa_CuxBi2Se3}.
The symmetries of the order parameter, along with Fermi surface topology, determine whether the superconducting state is topological; hence, identifying them is of significant importance~\cite{sato2017topological,sato2016majorana,tanaka2011symmetry}.
Moreover, non-unitary spin-triplet states realize a spin-polarized condensate, offering a platform for magneto-superconducting phenomena, such as a spin-polarized supercurrent~\cite{matsushita_SSSE}.

Our work continues recent efforts to use spin caloritronic phenomena, such as the SNE and the superconducing spin-Seebeck effect (SSSE), as sensitive probes of the spin structure of the condensate~\cite{matsuhita_SNE1,matsushita_SSSE}.
These studies employed the quasiclassical Keldysh (Eilenberger) theory, which, in its standard form, keeps only the leading-order terms in an expansion in powers of the small parameter $(k_{\rm F}\xi_0)^{-1}\sim T_{\mathrm c}/\epsilon_{\rm F}$, where $k_{\rm F}$ is the Fermi momentum, $\xi_0$ is the superconducting coherence length, $T_{\mathrm c}$ is the superconducting transition temperature, and $\epsilon_{\rm F}$ is the Fermi energy~\cite{eilenberger1968transformation,serene1983quasiclassical,rainer1995strong}.
Truncation at leading order omits the topological contributions arising from the Berry curvature, which are only included in higher-order terms~\cite{kobayashi_NTMR}. 
We therefore take a direct approach using the Bogoliubov-de Gennes (BdG) Hamiltonian below.

The rest of this paper is organized as follows.
In Sec.~\ref{sec: def_spincurrent}, we introduce the definition of the conserved spin current.
In Sec.~\ref{sec: SNC}, we derive the expression for the SNC.
The quasiparticle contribution to the SNC was originally derived in Ref.~\cite{xiao_conSC}; for completeness, we provide a self-contained derivation in Appendices~A--K (for readability, the derivation is divided into several steps).
As emphasized above, we complement that derivation by including the supercurrent contribution.
In Sec.~\ref{sec: model}, we present a simple model for spin-triplet superconductors.
Sections~\ref{sec: ISNE_helicalsc} and~\ref{sec: ISNE_nonunitarySC} form the core of this paper: we show that the $d$-vector induces the Berry curvature generated via the coupling between the spin and relative orbital motion of spin-triplet Cooper pairs, thereby giving rise to the intrinsic SNE.
Section~\ref{sec: ISNE_helicalsc} focuses on a time-reversal-invariant helical state, and Sec.~\ref{sec: ISNE_nonunitarySC} addresses the intrinsic SNE in nonunitary (time-reversal-symmetry-broken) superconductors. 
We conclude with a brief section placing our results in the context of recent work on topological superconductors.

Throughout this paper, we use the notation $\hat{A}$ for an operator and $A$ for its expectation value.
For simplicity, we set $e=k_{\rm B}=\hbar=1$ throughout the paper.

\section{Definitions of spin current}
\label{sec: def_spincurrent}
\label{sec: spincurrent}
The phase transition to a spin-triplet superconducting state spontaneously breaks not only the U(1) gauge symmetry but also spin conservation. When analyzing spin transport, we need to keep in mind that the net quasiparticle spin along a particular direction in a given volume changes not only through inflow/outflow of the spin current but also via spin precession.
Consequently, the spin torque density $\tau^\mu = d s^\mu / dt$ must appear in the spin continuity equation for quasiparticles:
\beq
\frac{\partial s^\mu}{\partial t}+{\bm \nabla}\cdot \bm J^{s^\mu}=\tau^{\mu}.
\eeq
The spin torque density on the right-hand side renders the definition of the spin current ambiguous.
The most commonly used definition--the so-called conventional spin current--is~\cite{sinova_SHE}
\beq
\label{eq: ConvSC}
\hat{\bm J}^{s^\mu} = \frac{\{\hat{s}^\mu, \bm \hat{v} \}}{2}\,.
\label{eq:Js_conv}
\eeq
Here the curly brackets $\{\hat{X}, \hat{Y}\}=\hat{X}\hat{Y}+\hat{Y}\hat{X}$ denote the anticommutator, $\hat{s}^\mu$ is the spin operator along the $\mu$ axis, and $\bm \hat{v}(\bm k)=\partial \hat{\mathcal{H}}(\bm k)/\partial \bm k$ is the quasiparticle velocity operator obtained from the BdG mean field Hamiltonian $\hat{\mathcal{H}}(\bm k)$.
In spin-conserved systems, Eq.~\eqref{eq:Js_conv} reduces to the difference between the charge currents of each spin projection.
However, when spin is not conserved, this definition is clearly flawed. 
For example, it yields a finite equilibrium expectation value for spin current even in spatially homogeneous noncentrosymmetric systems~\cite{Rashba_SC}.

A conserved spin current was developed to resolve such issues~\cite{shi_conSC,zhang_conSC,murakami_conSC,xiao_mott}. 
Following this approach, we express the spin torque density as the divergence of the spin torque dipole density $\bm J^{\tau^\mu}$
\beq
\label{eq: torquedipole}
\tau^{\mu}(\bm x) = -{\bm \nabla} \cdot \bm J^{\tau^\mu}.
\eeq
Moving the spin torque density to the left-hand side, we obtain the continuity equation of spin without a source term
\beq
\label{eq: continuity2}
\frac{\partial s^\mu}{\partial t} + {\bm \nabla} \cdot \left( \bm J^{s^\mu} + \bm J^{\tau^\mu} \right) = 0\,,
\eeq
which defines the conserved spin current as
\beq
\label{eq: ConsSC1}
\bm{\mathcal{J}}^{s^\mu} = \bm J^{s^\mu} + \bm J^{\tau^\mu}.
\eeq
At the operator level, the conserved spin current operator is expressed as
\beq
\label{eq: ConsSC2}
\bm{\hat{\mathcal{J}}}^{s^\mu}=\frac{d\left(\bm r\hat{s}^\mu\right)}{dt}=\hat{\bm J}^{s^\mu}+\frac{\left\{ \bm r,\hat{\tau}^{\mu}\right\}}{2},
\eeq
where $\hat{\tau}^{\mu}=d\hat{s}^\mu/dt$ represents the spin torque density operator.
Equation~\eqref{eq: ConsSC2} shows that, compared to the conventional spin current [Eq.~\eqref{eq:Js_conv}], the conserved spin current includes the additional term due to the spin torque dipole.

The conserved spin current satisfies the required properties, such as satisfying Onsager reciprocity relations and Mott relation~\cite{shi_conSC,zhang_conSC, murakami_conSC,xiao_mott}.
Importantly, its equilibrium value is a pure magnetization (circulating) current and therefore vanishes upon averaging over a cross section~\cite{xiao_conSC}:
\beq
\bm{\mathcal{J}}_{\rm leq}^{s^\mu} = \bm \nabla \times \bm M^{s^\mu}.
\eeq
This result was derived initially in electronic systems in Ref.~\cite{xiao_conSC} and has more recently been extended to localized spin systems~\cite{koyama}. 
For completeness, we provide a self-contained derivation in the Appendices~A-H. We employ this definition in our calculations.

\section{Spin Nernst conductivity}
\label{sec: SNC}
The SNE is part of the general linear response of superconductors to temperature gradients, in which the spin ($\mathcal{J}_i^{s^\mu}$) and charge ($\mathcal{J}_i$) currents are given by
\begin{subequations}
\beq
\label{eq: linearres_spincurrent}
\mathcal{J}_{i}^{s^\mu}&=&\alpha_{ij}^{s^\mu}(-\partial_{r_j}T)+D^{s^\mu}_{ij}(-\partial_{r_j}\theta),\\
\label{eq: linearres_chargecurrent}
\mathcal{J}_{i}&=&\alpha_{ij}(-\partial_{r_j}T)+D_{ij}(-\partial_{r_j}\theta).
\eeq
\label{eq:linearres}
\end{subequations}
Throughout the manuscript, we adopt the Einstein convention, where repeated indices are implicitly summed over.
Here, $-\partial_{r_j}T$ and $-\partial_{r_j}\theta$ represent the gradients of the temperature and the superconducting phase, respectively.
The tensors $D^{s^\mu}_{ij}$ and $D_{ij}$ denote the spin and charge superfluid weights, respectively.
Their expressions for our models are given below in Eq.~\eqref{eq:sup_dens}.

The first terms on the right-hand side of Eqs.~\eqref{eq: linearres_spincurrent} and \eqref{eq: linearres_chargecurrent} describe the direct quasiparticle response to temperature gradients.
For $i\neq j$, they read
\begin{subequations}
\begin{align}
\label{eq:alpha_smu}
&\alpha_{ij}^{s^\mu} = \frac{1}{V} \sum_{n}\sum_{\bm k} 
\left[s_n^\mu \left(\mathcal{B}_{ij}^{kk}\right)_n+\partial_{k_j}\left(d_{i}^{s^\mu}\right)_n\right] \mathcal{S}(\epsilon_{n\bm k}),\\
\label{eq:alpha_mu}
&\alpha_{ij} = \frac{1}{V} \sum_n \sum_{\bm k}  \left[\rho_n \left(\mathcal{B}_{ij}^{kk}\right)_n+\partial_{k_j}\left(d^{\rho}_{i}\right)_n\right] \mathcal{S}(\epsilon_{n\bm k}).
\end{align}
\label{eqs:alpha}
\end{subequations}
For completeness, we present a detailed derivation of these results in the Appendices~A--K, following Ref.~\cite{xiao_conSC}, in which Eqs.~\eqref{eqs:alpha} were originally derived. 

In Eq.~\eqref{eqs:alpha}, the expectation values of the spin and the quasiparticle charge are $s_n^\mu(\bm k)=\langle u_{n\bm k} | \hat{s}^\mu | u_{n\bm k} \rangle$ and $\rho_n(\bm k)=\langle u_{n\bm k} | \hat{\rho} | u_{n\bm k} \rangle$, where $\hat{\rho}$ is the charge operator.
The quantity $\left(\mathcal{B}_{ij}^{kk}(\bm k)\right)_n$ denotes the momentum space Berry curvature for the $n$-th band, defined as~\cite{xiao_berrycur}
\beq
\label{eq: Berrycurvature}
\left(\mathcal{B}_{ij}^{kk}\right)_n = -2\Im \sum_{m (\ne n)} \frac{\langle u_{n\bm k} | \hat{v}^i | u_{m\bm k} \rangle \langle u_{m\bm k} | \hat{v}^j | u_{n\bm k} \rangle}{(\epsilon_{n \bm k} - \epsilon_{m \bm k})^2},
\eeq
where $|u_{n\bm k}\rangle$ is the quasiparticle eigenstate with the energy $\epsilon_{n \bm k}$.
The quantities $\left(d_{i}^{s^\mu}\right)_n$ and $\left(d^{\rho}_{i}\right)_n$ represent the spin magnetic quadrupole moment and the charge dipole moment of quasiparticles, respectively~\cite{shitade_SHE}
\beq
\left(d_{i}^{s^\mu}\right)_n&=&\Re \sum_{m(\neq n)} (\mathcal{A}_i^k)_{nm}(s^\mu)_{mn},\\
\left(d^{\rho}_i\right)_{n}&=&\Re \sum_{m(\neq n)} (\mathcal{A}_i^k)_{nm}(\rho)_{mn},
\eeq
where the off-diagonal matrix elements of the spin and charge operators are $s_{mn}^\mu(\bm k)=\langle u_{m\bm k} | \hat{s}^\mu | u_{n\bm k} \rangle$ and $\rho_{mn}(\bm k)=\langle u_{m\bm k} | \hat{\rho} | u_{n\bm k} \rangle$ $(m\neq n)$.
Finally, $(\mathcal{A}_i^k)_{nm}=i\langle u_{m\bm k} | \partial_{k_i} | u_{n\bm k} \rangle$ denotes the Berry connection in momentum space, and $\mathcal{S}(\epsilon_{n \bm k})$ stands for the state-resolved entropy density
\beq
T_{\rm eq} \mathcal{S}(\epsilon_{n \bm k}) = \epsilon_{n \bm k}f(\epsilon_{n \bm k}) + T_{\rm eq} \ln \left(1 + e^{-\epsilon_{n \bm k}/T_{\rm eq}} \right).
\label{eq:entropy_density}
\eeq
Here $T_{\rm eq}$ is the (spatially homogeneous) equilibrium temperature and $f(\epsilon)=1/(1+e^{\epsilon/T_{\rm eq}})$ is the Fermi-Dirac distribution function, where $\epsilon$ is measured relative to the chemical potential.

It is important to compare Eq.~\eqref{eqs:alpha} with the corresponding result obtained in the Kubo-Luttinger formalism using the conventional spin current.~\cite{luttinger1964theory,smrcka1977transport,cooper_thermoelectric}.
In this approach, the SNC ($i\neq j$) is given by~\cite{fujimoto_SNE}
\beq
\label{eq: SNE_convSC}
\beta_{ij}^{s^\mu} &=& J_i^{s^\mu}/(-\partial_{r_{j}} T)\nn\\
&=& \frac{1}{V} \sum_n \sum_{\bm k} \left(\mathcal{B}_{ij}^{kks^\mu}\right)_n \mathcal{S}(\epsilon_{n \bm k})\,,
\eeq
where the spin Berry curvature is defined by~\cite{Lau_SHE,taguchi_SHE,matsushita_SNE2}
\beq
\label{spin_Berrycur}
\left(\mathcal{B}_{ij}^{kks^\mu}\right)_n = -2\Im \sum_{m (\ne n)} \frac{\langle u_{n\bm k} | \hat{v}^i | u_{m\bm k} \rangle \langle u_{m\bm k} | \hat{J}_j^{s^\mu} | u_{n\bm k} \rangle}{(\epsilon_{n \bm k} - \epsilon_{m \bm k})^2}.
\eeq
In contrast to Eq.~\eqref{eq:alpha_smu}, the spin magnetic quadrupole moment contribution is absent from Eq.~\eqref{eq: SNE_convSC}; consequently, $\beta_{ij}^{s^\mu}$ is entirely determined by the spin Berry curvature.
We emphasize that the spin Berry curvature does not generally coincide with the momentum space Berry curvature, except in spin-conserved systems, because the conventional spin current operator is not proportional to the velocity operator.
This crucial distinction is illustrated in Sec.~\ref {sec: ISNE_helicalsc}, where we show that the two approaches yield qualitatively different results for the spin-triplet superconducting states.

The second terms in Eq.~\eqref{eq:linearres} describe the thermoelectric spin and charge supercurrents induced by temperature gradients.
Their presence is required to cancel magnetic fields induced by the quasiparticle thermoelectric charge current in the bulk of the superconductor.
More specifically, as quasiparticles carry both electric charge and entropy, their motion in response to temperature gradients gives rise to the thermoelectric charge current.
This quasiparticle thermoelectric charge current is described by the first term in Eq.~\eqref{eq: linearres_chargecurrent}.
However, in the superconducting state, such a charge current can exist only within the charge imbalance length from surfaces~\cite{tinkham_SC, wakamura2015quasiparticle, takahashi2011spin}.
At longer length scales, the thermoelectric charge current is screened by the Meissner effect.
For simplicity, we neglect the surface effects and assume a complete magnetic flux exclusion, i.e., $\bm{B} = 0$.
According to the Maxwell-Amp\`ere law, this condition implies vanishing bulk charge current, $\bm{\mathcal{J}} \propto\bm{\nabla} \times \bm{B} = 0$.
Under this assumption, the superconducting phase gradient induced by temperature gradients is given by [see Eq.~\eqref{eq:linearres}]
\beq
\label{eq: vectorpotential}
(-\partial_{r_l}\theta)&=&-(D^{-1})_{lm}\alpha_{mj}(-\partial_{r_j}T).
\eeq
Substituting Eq.~\eqref{eq: vectorpotential} into Eq.~\eqref{eq: linearres_spincurrent}, we obtain the expression for the SNC
\beq
\label{eq: SNC}
\tilde{\alpha}_{ij}^{s^\mu}
=\mathcal{J}_{i}^{s^\mu}/(-\partial_{r_j}T)
=\alpha_{ij}^{s^\mu}-D^{s^\mu}_{il}(D^{-1})_{lm}\alpha_{mj}.
\eeq
The last term in Eq.~\eqref{eq: SNC} is unique to superconductors.  
Such a contribution is absent in normal metals, where the spin Hall effect induced by the Seebeck voltage modifies the spin Nernst current~\cite{fujimoto_SNE, matsushita_SNE2, tauber_SNE}.

\begin{figure}[t]
\includegraphics[width=70mm]{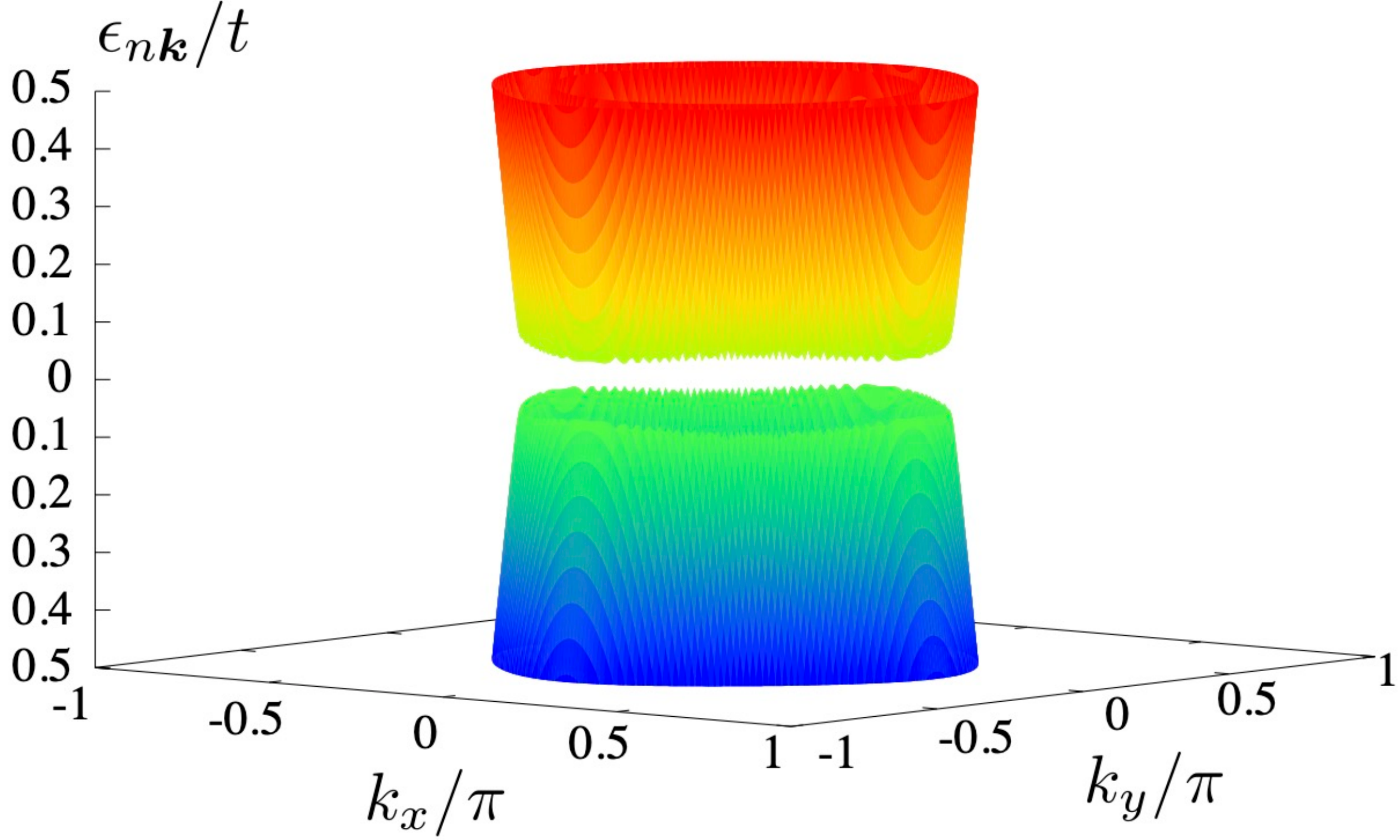}
\caption{
Quasiparticle spectrum in the helical superconducting state with the $d$-vector in Eq.~\eqref{eq: dvector_helicalstate}.
The parameters are set to $t=1.25$ meV, with $(\mu/t,\Delta_0(T_{\rm eq})/t)=(-2,0.05)$.
}
\label{fig1-1}
\end{figure}

\begin{figure}[b]
\includegraphics[width=80mm]{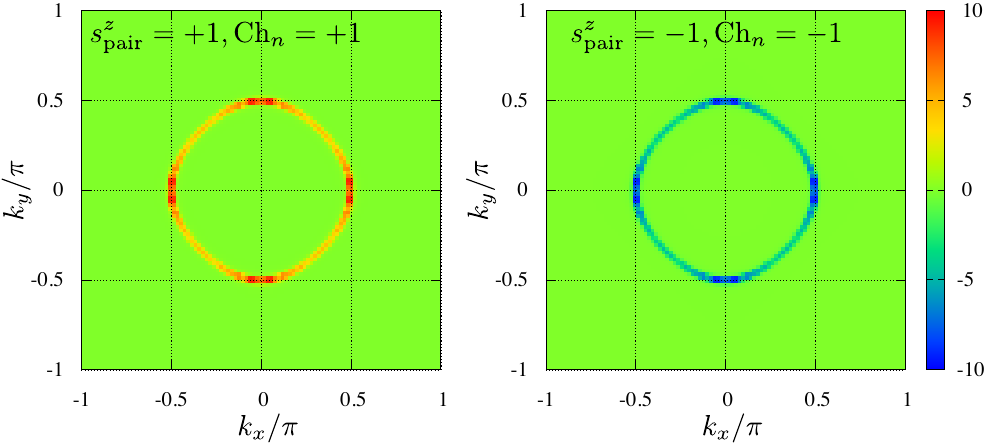}
\caption{
Berry curvature, along with the spin of Cooper pairs $s_{\rm pair}^z=\pm 1$ and the band-resolved Chern numbers ${\rm Ch}_n$, for the negative-energy quasiparticle states in the helical superconductor.
The parameters are set to $t=1.25$ meV, with $(\mu/t,\Delta_0(T_{\rm eq})/t)=(-2,0.05)$.
}
\label{fig1-2}
\end{figure}

\section{Model Hamiltonian}
\label{sec: model}
Hereafter, we consider two-dimensional spin-triplet superconductors described by the BdG Hamiltonian in spin and Nambu (particle-hole) space~\cite{sigrist_SC}
\begin{subequations}
\begin{eqnarray}
\hat{\mathcal{H}}(\bm k)  &=& \begin{pmatrix}
    \xi(\bm k)\hat{s}^0 & \hat{\Delta}(\bm k)\\
    \hat{\Delta}^\dagger(\bm k) & -\xi(\bm k)\hat{s}^0
\end{pmatrix}\,,
\\
\hat{\Delta}(\bm k) &=&i(\bm{d}(\bm{k})\cdot{\hat{\bm s}})\hat{s}^y\nn\\
&=&
\begin{pmatrix}
    -d_x(\bm k)+i d_y(\bm k) & d_z(\bm k)\\
    d_z(\bm k) & d_x(\bm k)+i d_y(\bm k)
\end{pmatrix}\,.
\end{eqnarray}
\label{eq: General_Hamiltonian}
\end{subequations}
We consider the tight-binding model on a square lattice with the normal-state dispersion $\xi(\bm k) = -2t (\cos k_xa + \cos k_ya) - \mu$, where $\bm k$ is the crystalline momentum, $a$ is a lattice constant, and $t$ is the nearest-neighbor hopping amplitude.
Here, the normal-state energy is measured from the chemical potential $\mu$.
$\hat{s}^0$ is the 2$\times$2 identity matrix and $\hat{\bm s}=(\hat{s}^x,\hat{s}^y,\hat{s}^z)$ is the vector of Pauli matrices in spin space.
The $d$-vector $\bm{d}(\bm{k}) = (d_x(\bm{k}), d_y(\bm{k}), d_z(\bm{k}))$ characterizes spin-triplet (total spin $s_{\rm pair}=1$) Cooper pairs. 
Antisymmetry under fermion exchange dictates $\bm{d}(\bm{k}) = -\bm{d}(-\bm{k})$.
The spin-polarization of the condensate at each momentum is given by $\delta {\bm s}_{\rm pair}(\bm k)=i\bm{d}(\bm{k})\times \bm{d}^\ast(\bm{k})$.
A spin-triplet superconducting state is called the unitary state if $\delta {\bm s}_{\rm pair}(\bm k)=0$ for all $\bm k$; otherwise, it is called the nonunitary state~\cite{sigrist_SC}.

For this model, the spin and charge superfluid weights read~\cite{hirobe2025anomalous}
\begin{subequations}
\begin{eqnarray}
D_{ij}^{s_z}&=&\frac{1}{V}\sum_{\bm k}\sum_{n,m}\frac{f(\epsilon_{n\bm k})-f(\epsilon_{m\bm k})}{\epsilon_{n\bm k}-\epsilon_{m\bm k}}\nn\\
&\times&\langle u_{n\bm k}|\hat{J}_i^{s_z}|u_{m\bm k}\rangle \langle u_{m\bm k}|\hat{J}_j|u_{n\bm k}\rangle \nn\\
&+&\frac{1}{V}\sum_{\bm k}\sum_{n}f(\epsilon_{n\bm k})\langle u_{n\bm k}|\partial_{k_i}\partial_{k_j}\hat{\mathcal{H}}_{\rm N}(\bm k)\hat{s}_z|u_{n\bm k}\rangle,\\
D_{ij}&=&\frac{1}{V}\sum_{\bm k}\sum_{n,m}\frac{f(\epsilon_{n\bm k})-f(\epsilon_{m\bm k})}{\epsilon_{n\bm k}-\epsilon_{m\bm k}}\nn\\
&\times&\langle u_{n\bm k}|\hat{J}_i|u_{m\bm k}\rangle \langle u_{m\bm k}|\hat{J}_j|u_{n\bm k}\rangle \nn\\
&+&\frac{1}{V}\sum_{\bm k}\sum_{n}f(\epsilon_{n\bm k})\langle u_{n\bm k}|\partial_{k_i}\partial_{k_j}\hat{\mathcal{H}}_{\rm N}(\bm k)|u_{n\bm k}\rangle.
\end{eqnarray}
\label{eq:sup_dens}
\end{subequations}
Here, $\hat{\mathcal{H}}_{\rm N}(\bm k)$ is the normal-state Hamiltonian in the Nambu space, expressed as
\beq
\hat{\mathcal{H}}_{\rm N}(\bm k)  &=& \begin{pmatrix}
    \xi(\bm k)\hat{s}^0 & 0_{2\times2}\\
    0_{2\times2} & -\xi(\bm k)\hat{s}^0
\end{pmatrix},
\eeq
where $0_{2\times2}$ is the $2\times 2$ zero matrix.

In the present two-dimensional single-orbital model without explicit antisymmetric spin-orbit coupling (SOC), the spin and charge superfluid-weight tensors have diagonal form, \(D_{xy}=D_{yx}=0\), and similarly for \(D_{ij}^{s^\mu}\).
Therefore, for the response \(\tilde{\alpha}_{xy}^{s_z}\), the supercurrent correction reduces to
\beq
\tilde{\alpha}_{xy}^{s_z}-\alpha_{xy}^{s_z}=-\frac{D_{xx}^{s_z}}{D_{xx}}\alpha_{xy}
\eeq
In lower-symmetry multiorbital systems or systems with antisymmetric SOC, off-diagonal components of the superfluid weight may appear.

The BdG Hamiltonian in Eq.~\eqref{eq: General_Hamiltonian} does not explicitly include antisymmetric SOC.
Nevertheless, the effect of SOC is phenomenologically taken into account through the assumed form of the \(d\)-vector.
This treatment is justified in the limit where the SOC energy scale \(E_{\rm SOC}\) is much larger than the superconducting transition temperature \(T_{\mathrm c}\).
This condition is expected to hold in uranium-based superconductors, where \(E_{\rm SOC}\sim 100\) K and \(T_{\mathrm c}\sim 1\) K, and which are promising candidates for spin-triplet superconductivity.
In this limit, SOC pins the \(\bm d\)-vector to specific crystalline directions~\cite{sigrist_SC}, and this pinning is incorporated by choosing the corresponding \(d\)-vector structure.
If inversion symmetry is broken locally or globally, the associated antisymmetric SOC can induce the SNE already in the normal state~\cite{sheng2017,bose2018,bose2019,cheng2008,chuu2010,dyrdal2016,dyrdal2016a,tauber_SNE,akera_SNE,fujimoto_SNE,matsushita_SNE2}.
However, this normal-state contribution peaks near $T_{\rm eq}\sim E_{\rm SOC}$ and rapidly becomes negligible for $T_{\rm eq}\ll E_{\rm SOC}$, the regime of interest here.

The integrand of the SNC in Eq.~\eqref{eq:alpha_smu} includes the Berry curvature in momentum space. 
In the spin-triplet superconducting state, the superconducting order parameter generates the Berry curvature, and the quasiparticle bands inherit its symmetry and topology; consequently, the $\bm{d}$-vector contribution to the intrinsic SNE becomes significant in topological phases in spin-triplet superconductors.
In the following, we therefore evaluate the SNC in spin-triplet topological superconductors both with and without time-reversal symmetry.

\section{Intrinsic Spin Nernst effect in helical superconductors}
\label{sec: ISNE_helicalsc}
We begin by considering a two-dimensional helical superconductor described by
\beq
\label{eq: dvector_helicalstate}
\bm d(\bm k)=\Delta_0(T_{\rm eq})(\sin k_xa,\sin k_ya,0)\,,
\eeq
and model the temperature dependence of the gap amplitude as~\cite{tinkham_SC}
\beq
\label{gap}
\Delta_0(T_{\rm eq})=1.765T_{\mathrm c}\tanh \left(1.74 \sqrt{\frac{T_{\mathrm c}}{T_{\rm eq}}-1}\right).
\eeq
Figure~\ref{fig1-1} shows that the helical superconducting order opens an energy gap at the chemical potential in the quasiparticle spectrum.
Because the helical state does not exhibit the spin-polarization $\delta {\bm s}_{\rm pair}(\bm k)=0$, the quasiparticle states have twofold spin degeneracy~\cite{sigrist_SC}.

If we write the helical order parameter in the spin space explicitly, 
\begin{equation}
    \hat{\Delta}(\bm k) =\Delta_0
\begin{pmatrix}
    -\sin k_xa+i\sin k_ya & 0\\
    0& \sin k_xa+i\sin k_ya,
\end{pmatrix}\,,
\label{eq: Hamiltonian_helical}
\end{equation}
it becomes clear that it is characterized by a spin-dependent phase winding around the $\Gamma$-point in momentum space, $\Delta_0(T_{\rm eq})(\mp \sin k_x+i\sin k_y)\simeq \mp \Delta_0(T_{\rm eq})e^{\mp i\phi_k}$, where $\phi_k=\tan^{-1}(k_y/k_x)$.
Thus, helical superconductors can be viewed 
as a superposition of two chiral superconductors with opposite phase winding in the two spin sectors ($s_{\rm pair}^z=\pm1$).
The chiral superconducting order in each spin sector realizes the class D topological superconductors characterized by the nonzero Chern number over a wide range of parameters~\cite{sato2017topological,sato2016majorana,tanaka2011symmetry}.
Recall that quantum spin Hall insulators can be regarded as two time-reversal copies of Chern insulators~\cite{kane_QSHE, kane_z2TI}.
Analogously, the helical superconducting order realizes the class DIII topological superconductors as a superposition of chiral states with opposite Chern numbers.
Figure~\ref{fig1-2} (b) shows the momentum space Berry curvature for the spin-degenerate states, indicating that each spin-degenerate state possesses an opposite value of the state-resolved Chern number, defined as
\beq
{\rm Ch}_n=\int \frac{d\bm k}{2\pi} \left(\mathcal{B}_{xy}^{kk}\right)_n,
\eeq
which renders the Kane-Mele $\mathbb{Z}_2$ topology nontrivial, thereby realizing the class DIII topological phase~\cite{kane_QSHE, kane_z2TI, sato2016majorana, sato2017topological,tanaka2011symmetry}.

\begin{figure}[b]
\includegraphics[width=80mm]{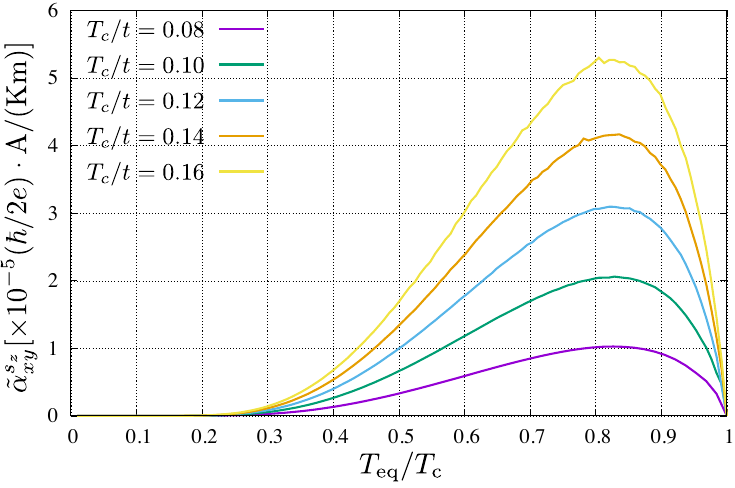}
\caption{
Temperature dependence of $\tilde{\alpha}_{xy}^{s^z}$ with several values of $T_{\mathrm c}/t$ in the helical superconductor.
The parameters are set to $t=1.25$ meV and $\mu=-2.5$ meV (bandwidth $E_{\mathrm B}=8t=10\;\mathrm{meV}$; equivalently $E_{\mathrm B}/k_{\mathrm B}=116\;\mathrm K$).
The plotted values are obtained by converting the two-dimensional SNC to the three-dimensional bulk value by dividing by the film thickness $d=2\mu\mathrm{m}$.
}
\label{fig2}
\end{figure}

\begin{figure}[t]
\includegraphics[width=80mm]{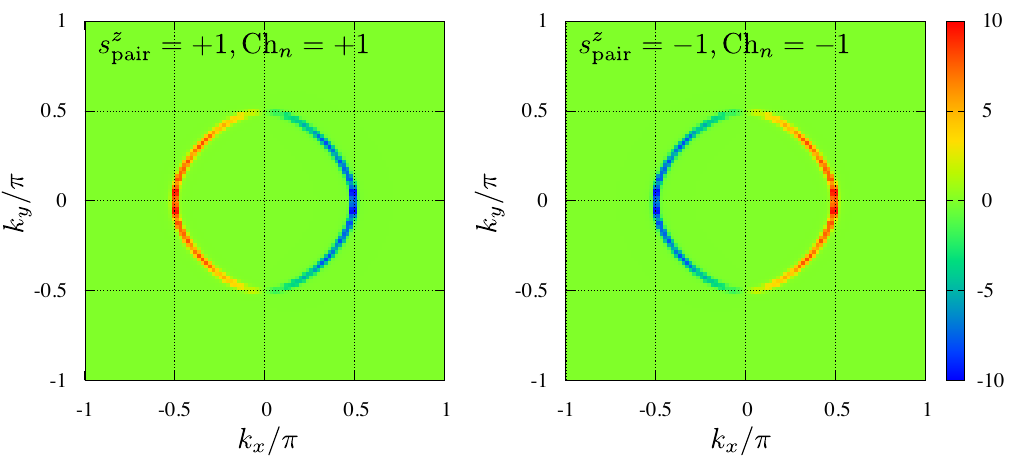}
\caption{
The spin magnetic quadrupole moment $(d_x^{s^z}(\bm k))_n$ for the degenerate occupied states in the helical superconductor.
The parameters are set to $t=1.25$ meV, with $(\mu/t,\Delta_0(T_{\rm eq})/t)=(-2,0.05)$.
}
\label{fig3}
\end{figure}

The SNC is constrained by the mirror-reflection symmetries that remain once strong SOC pins the direction of the $\bm d$-vector~\cite{matsushita_SSSE, Seemann2015}.
In the helical state, the mirror reflection symmetries about the $yz$ and $zx$ planes are preserved, leading to the following constraints:
\beq
\tilde{\alpha}_{xy}^{s^x}=\tilde{\alpha}_{yx}^{s^x}=\tilde{\alpha}_{xy}^{s^y}=\tilde{\alpha}_{yx}^{s^y}=0.
\eeq
Hence, $\tilde{\alpha}_{xy}^{s^z} = -\tilde{\alpha}_{yx}^{s^z}$ remains as the only symmetry-allowed nonzero component of the SNC.
We therefore focus on the SNE with the $z$-polarized spin current.
Physically, this stems from the fact that in Eq.~\eqref{eq: Hamiltonian_helical} the order parameter is nonzero only in the equal-spin pairing sector ($s_{\rm pair}^z=\pm1$).

Figure~\ref{fig2} shows the temperature dependence of $\tilde{\alpha}_{xy}^{s^z}$ for several values of $T_{\mathrm c}/t$ and clearly demonstrates a finite $d$-vector contribution to the intrinsic SNE.
This contribution emerges and grows below $T_{\mathrm c}$ due to the evolution of the spin-triplet order parameter.
However, because this state is fully gapped in two dimensions, the SNC is exponentially suppressed at low temperatures ($T_{\rm eq}/T_{\mathrm c} \ll 1$).
The helical state preserves time-reversal symmetry, and the spin superfluid weight vanishes due to the identical quasiparticle spectra in the $s_{\rm pair}^z=\pm 1$ subspaces.
According to Eq.~\eqref{eq: SNC}, the supercurrent contribution is absent, so that $\tilde{\alpha}_{xy}^{s^z} = \alpha_{xy}^{s^z}$.
Furthermore, Fig.~\ref{fig3} shows that the contribution of the spin quadrupole moment to Eq.~\eqref{eq:alpha_smu}, $\partial_{k_y}\left(d_{x}^{s^\mu}(\bm k)\right)_n$, vanishes after the momentum integration.
As a result, in helical superconductors, the SNC is entirely determined by the Berry curvature generated by the $\bm{d}$-vector and the quasiparticle spectrum.

It is instructive to compare our result on $\bm{\hat{\mathcal{J}}}^{s^\mu}$ with that obtained by using the conventional spin current.
For the Hamiltonian in Eq.~\eqref{eq: General_Hamiltonian}, the conventional spin current operator reads
\beq
\label{eq: conventionalspincurrent}
\hat{\bm J}^{s^\mu}&=&\frac{\partial \xi(\bm k)}{\partial \bm k}\hat{s}^\mu \hat{\sigma}^0-\Re \left[\frac{\partial d_\mu(\bm k)}{\partial \bm k} \right]\hat{s}^y \hat{\sigma}^y-\Im \left[\frac{\partial d_\mu(\bm k)}{\partial \bm k} \right]\hat{s}^y \hat{\sigma}^x,\nonumber\\
\eeq
where $\hat{\sigma}^0$ and $\hat{\sigma}^j$ ($j = x, y, z$) denote the identity and Pauli matrices in Nambu (particle-hole) space.
The $\mu$-component of the $d$-vector preserves $\hat{s}^\mu$ for quasiparticles and thus contributes to $\hat{\bm J}^{s^\mu}$.
The other components do not preserve $\hat{s}^\mu$ and they do not appear in $\hat{\bm J}^{s^\mu}$.
As a result, the $d$-vector contribution to $\beta_{ij}^{s^\mu}$ vanishes in many cases.
Substituting the $d$-vector in the helical state into Eq.~\eqref{eq: conventionalspincurrent}, we obtain 
\beq
\label{eq: conventionalspincurrent_helicalstate}
\hat{\bm J}^{s^z}&=&\frac{\partial \xi(\bm k)}{\partial \bm k}\hat{s}^z \hat{\sigma}^0,
\eeq
which leads to a vanishing spin Berry curvature $\left(\mathcal{B}_{ij}^{kks^\mu}(\bm k)\right)_n=0$ and hence $\beta_{xy}^{s^z}=0$.

The expressions in Eq.~\eqref{eqs:alpha} contain the momentum space Berry curvature as a consequence of including the spin torque dipole [see Appendix~\ref{sec: derivation SNC} for details]. Because the superconducting order parameter matrix in Eq.~\eqref{eq: Hamiltonian_helical} is diagonal in spin space, the $z$-spin component of Cooper pairs ($s^z_{\rm pair}$) is conserved in the helical state.
In contrast, quasiparticles are an unequal-weight mixture of electrons and holes carrying spin angular momenta, so $\hat{s}^z$ is not a good quantum number for them.
Put simply, the spin structure of the condensate does not directly translate into the conventional spin current.
Accounting for the spin torque dipole, $\tau^{\mu}({\bm x})$, is therefore crucial in this and similar cases, yielding a non-vanishing SNC.

We contrast the helical superconductor considered above with the case where the $d$-vector has a single component, $\bm d(\bm k) \propto (0,0,(k_x \pm ik_y)^n)\; (n\in \mathbb{Z}, |n|\geq 1)$, corresponding to the spin-projection $s_{\rm pair}^z=0$ of Cooper pairs.
The superconducting state with such a $d$-vector, referred to as chiral, preserves the twofold spin-rotational symmetry about the $x$-axis in spin space, i.e., the invariance under a $\pi$-rotation acting only on spin while leaving the crystalline momentum unchanged~\cite{matsushita_SSSE}.
We emphasize that this does not imply the absence of SOC; rather, it reflects the {\em residual} symmetry remaining once SOC pins the direction of the $\bm d$-vector. 
As a result, the contributions to the spin current from different spin sectors cancel out in the absence of an external magnetic field.
Applying a magnetic field along the $z$-axis breaks this twofold spin-rotational symmetry by shifting the quasiparticle occupation.
In this situation, $\hat{s}_z$ is conserved and the spin torque vanishes, so the conventional spin current formulation suffices.
This situation may be realized in the B phase of UPt$_3$~\cite{yanaseupt3,nomoto_UPt3}.

\begin{figure}[t]
\includegraphics[width=80mm]{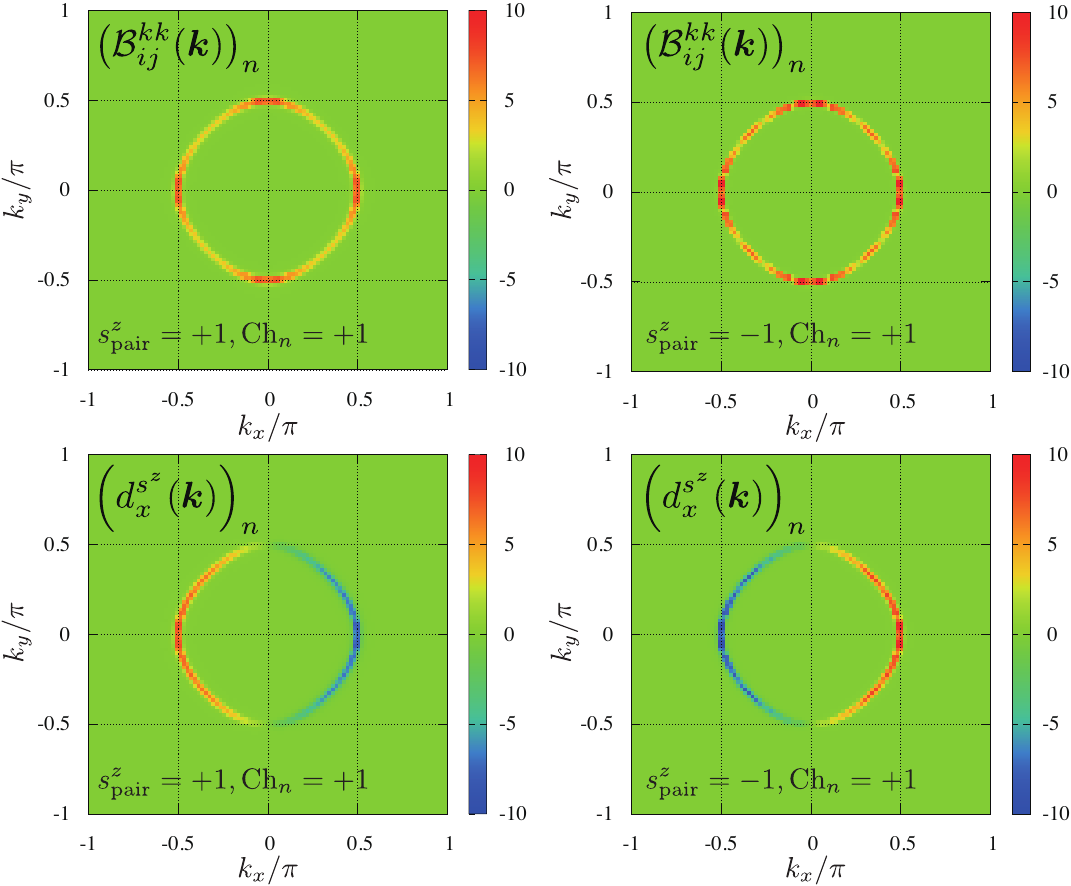}
\caption{
Upper panels: Berry curvature and band-resolved Chern numbers ${\rm Ch}_n$ for the negative-energy quasiparticle states in the nonunitary superconductor, Eq.~\eqref{eq:OP_nonun}.
Lower panels: Spin magnetic quadrupole moment and band-resolved Chern numbers ${\rm Ch}_n$ for the negative-energy quasiparticle states in the nonunitary superconductor.
The parameters are set to $t=1.25$ meV, with $(\mu/t,\Delta_0(T_{\rm eq})/t,\eta)=(-2,0.05,0.4)$.
}
\label{fig5}
\end{figure}

\begin{figure*}[t]
\includegraphics[width=160mm]{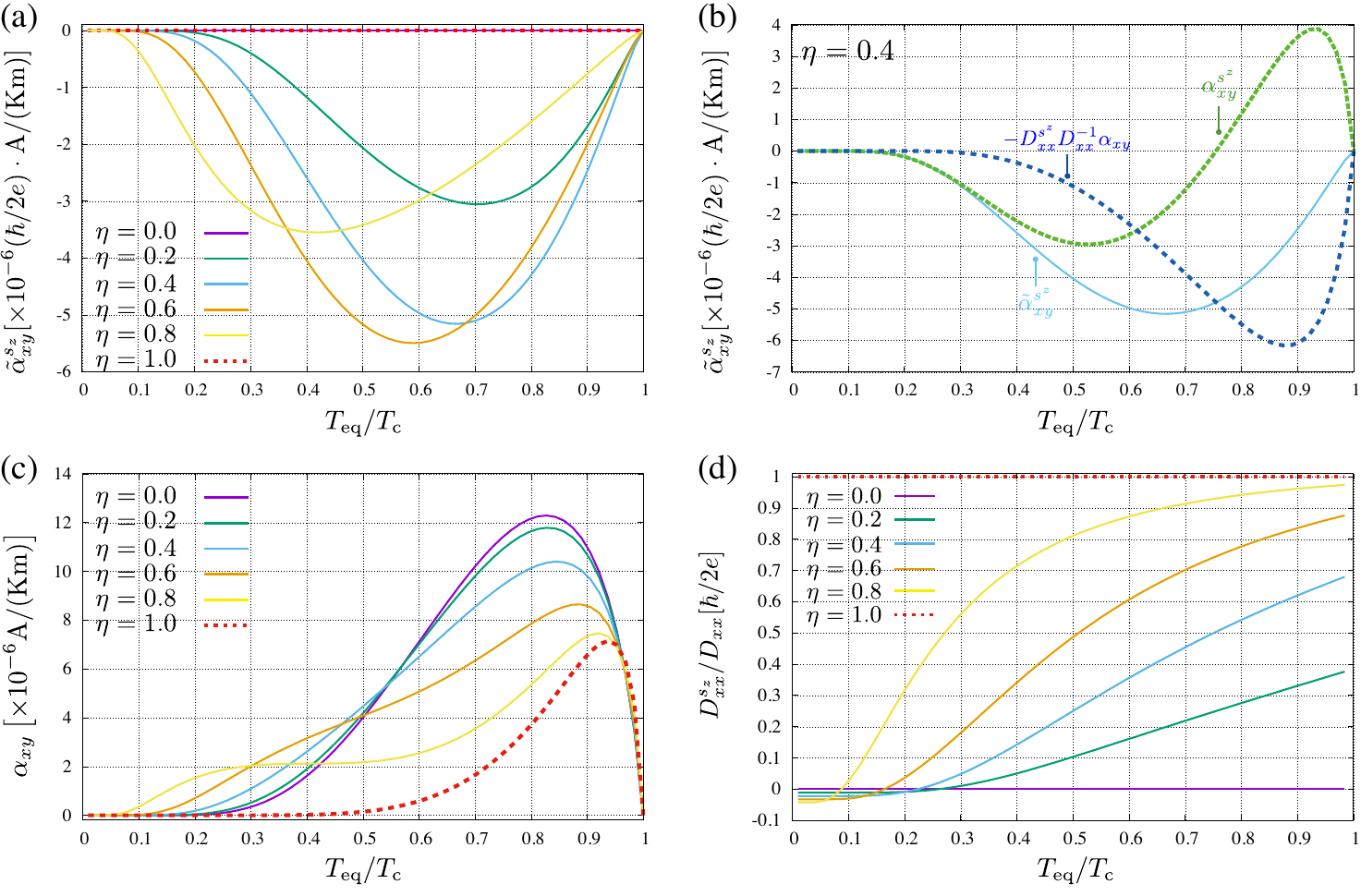}
\caption{
Temperature dependencies of (a-b) the SNC, (c) $\alpha_{xy}$, and (d) $D_{xx}^{s^z}/D_{xx}$ in nonunitary superconductors are shown.
Panels (a) and (c-d) present calculated results for several values of $\eta$.
Panel (b) illustrates individual contributions to the SNC for $\eta=0.4$.
The parameters are set to $t=1.25$ meV (bandwidth $E_{\mathrm B}=8t=10\;\mathrm{meV}$), with $(\mu/t, T_{\mathrm c}/t) = (-2, 0.05)$.
 }
\label{fig4}
\end{figure*}

\section{Intrinsic Spin Nernst effect in nonunitary superconductors}
\label{sec: ISNE_nonunitarySC}
Next, we consider a two-dimensional nonunitary superconductor described by
\beq
\bm d(\bm k)=\Delta_0(T_{\rm eq})\left(\sin k_xa+i \sin k_ya,\eta \left(\sin k_ya-i \sin k_xa\right),0\right),\nn\\
\label{eq:OP_nonun}
\eeq
where $\Delta_0(T_{\rm eq})$ is modeled by Eq.~\eqref{gap}.
The BdG Hamiltonian is still given by Eq.~\eqref{eq: General_Hamiltonian}, but the order parameter matrix in spin space now reads
\begin{eqnarray}
    \hat{\Delta}(\bm k)
    =
\begin{pmatrix}
    \Delta_{\uparrow \uparrow}(\sin k_xa+i\sin k_ya) & 0\\
    0& \Delta_{\downarrow \downarrow}(\sin k_xa+i\sin k_ya)
\end{pmatrix}\,,\nn\\
\label{eq: Hamiltonian_nonun}
\end{eqnarray}
with $\Delta_{\uparrow \uparrow}(T_{\rm eq})=\Delta_0(T_{\rm eq})(-1+\eta)$ and $\Delta_{\downarrow \downarrow}(T_{\rm eq})=\Delta_0(T_{\rm eq})(1+\eta)$.
A finite $\eta$ makes the gap amplitudes spin dependent, realizing a nonunitary superconducting state with the $z$ spin-polarization in the condensate. 
For $\eta > 0$, the gap amplitude in the $s^z_{\rm pair}=-1$ sector is larger than that in the $s^z_{\rm pair}=+1$ sector,
\beq
|\Delta_{\downarrow \downarrow}(T_{\rm eq})|>|\Delta_{\uparrow \uparrow}(T_{\rm eq})|.
\eeq
We therefore refer to the $s^z_{\rm pair}=-1$ and $s^z_{\rm pair}=+1$ sectors as the majority- and minority-spin sectors of the condensate, respectively.
Here, the terms majority and minority refer solely to the relative magnitudes of the superconducting gaps and not to the spin-resolved density of states in the normal state, which is identical for the two spin species in our model.
Such a nonunitary state may be realized in ferromagnetic superconductors such as UCoGe, where a spin-polarized condensate is stabilized by ferromagnetic moments coexisting with superconductivity~\cite{mineev_FMSC,hattori_UCoGe}, and the normal state DOS may be different for the two spin projections. 

For this order parameter, the $d$-vector exhibits a phase winding in momentum space, $\bm{d}(\bm{k}) \propto e^{i\phi_k}$, and the condensate acquires a polarized orbital angular momentum along the $z$-axis.
In contrast to the helical state considered above, the winding number is the same in both spin sectors, thereby realizing a class D topological phase~\cite{sato2016majorana,sato2017topological,tanaka2011symmetry}.
Figure~\ref{fig5} shows that the negative-energy quasiparticle bands share the same Chern number.

The twofold rotational symmetry, together with the mirror reflection symmetry, restricts the SNC tensor~\cite{matsushita_SSSE, Seemann2015}.
The nonunitary state above preserves the twofold rotational symmetry about the $z$-axis.
This symmetry prohibits the SNE with the $x$ and $y$ spin-polarization
\beq
\tilde{\alpha}_{xy}^{s^x}=\tilde{\alpha}_{yx}^{s^x}=\tilde{\alpha}_{xy}^{s^y}=\tilde{\alpha}_{yx}^{s^y}=0.
\eeq
We thus focus on the SNE with the $z$-polarized spin current.
Figure~\ref{fig5} also shows that $\partial_{k_y}\left(d_{x}^{s^\mu}(\bm k)\right)_n$ vanishes after integrating over the momentum.
As in the helical superconductor, the SNC is entirely determined by the Berry curvature generated by the $\bm{d}$-vector and the quasiparticle spectrum.

Figure~\ref{fig4} (a) shows the temperature dependence of the SNC in the nonunitary superconductor.
Note that the SNC vanishes both at $\eta = 0$ and at $\eta = 1$.
At $\eta = 0$, the superconducting state becomes unitary ($\delta \bm {s}_{\rm pair}(\bm k)=0$), where Cooper pairs are no longer spin-polarized for any $\bm k$.
Crucially, the pairing is identical in the spin-up and the spin-down channels [see Eq.~\eqref{eq: Hamiltonian_nonun}], corresponding to a single component $d$-vector $\bm d(\bm k)=(d_x, 0, 0)$.
This unitary state acquires an emergent twofold spin rotational symmetry about the $x$-axis, which leads to $\tilde{\alpha}_{xy}^{s^z} = 0$.
Hence, inequivalence between the pairing for the two spin projections is essential for the intrinsic SNE in spin-triplet superconductors.

For $\eta = 1$, the spin-polarization becomes complete, i.e., $\Delta_{\uparrow\uparrow}(T_{\rm eq})=0$.
In this limit, the spin current becomes proportional to the charge current, so the screening of the thermoelectric charge current also cancels the spin current.

Figure~\ref{fig4} (b) clearly shows that for intermediate values of the parameter $\eta$, the thermoelectric supercurrent contributes substantially to the SNC.
Both the quasiparticle charge current response [Eq.~\eqref{eq:alpha_mu}] and the spin superfluid density become finite in nonunitary superconductors, as shown in Fig.~\ref{fig4} (c,d).
The magnitudes of the individual contributions are comparable, indicating that the thermoelectric spin supercurrent must be taken into account for a proper evaluation of the SNE in nonunitary superconductors.

The quasiparticle contribution to the SNC changes sign as a function of temperature [Fig.~\ref{fig4}(b)].
This sign reversal stems from the spin-dependent gap amplitudes in the nonunitary superconducting state.
Because the two equal-spin pairing sectors contribute to $\alpha_{xy}^{s_z}$ with opposite signs, the sign of the total quasiparticle contribution is determined by which spin sector dominates the response.

Just below $T_{\mathrm c}$, where $\Delta_0(T_{\rm eq})/T_{\rm eq}\lesssim 1$, the superconducting gaps are comparable to or smaller than the thermal broadening. Consequently, the entropy-density factor in Eq.~\eqref{eq:entropy_density} is nearly identical in the two spin sectors. In contrast, the magnitude of the momentum-space factor entering Eq.~\eqref{eq:alpha_smu} is larger in the majority-spin sector because of its larger gap amplitude. The quasiparticle contribution near $T_{\mathrm c}$ is therefore dominated by the majority-spin sector.

The situation is reversed at low temperatures. Our model describes a fully gapped superconducting state, and the thermally excited quasiparticle population in each spin sector is therefore exponentially suppressed when $T_{\rm eq}/\Delta_0(T_{\rm eq})\ll 1$. Because the two spin sectors have different gap amplitudes, their activation energies are also different. The contributions from the $s^z_{\rm pair}=-1$ and $s^z_{\rm pair}=+1$ sectors are suppressed as $\exp[-|\Delta_{\downarrow\downarrow}(T_{\rm eq})|/T_{\rm eq}]$ and $\exp[-|\Delta_{\uparrow\uparrow}(T_{\rm eq})|/T_{\rm eq}]$, respectively. We denote the majority- and minority-spin contributions to $\alpha_{xy}^{s^z}$ by $\alpha_{xy}^{s^z({\rm maj})}$ and $\alpha_{xy}^{s^z({\rm min})}$. Their ratio behaves as
\beq
\left|
\frac{\alpha_{xy}^{s^z({\rm maj})}}
{\alpha_{xy}^{s^z({\rm min})}}
\right|
\sim
\exp\left[
-\frac{
|\Delta_{\downarrow\downarrow}(T_{\rm eq})|
-
|\Delta_{\uparrow\uparrow}(T_{\rm eq})|
}{T_{\rm eq}}
\right],
\eeq
up to algebraic prefactors. Thus, as the temperature is lowered, the contribution from the larger-gap $s^z_{\rm pair}=-1$ sector is suppressed more rapidly, whereas the contribution from the smaller-gap $s^z_{\rm pair}=+1$ sector remains thermally active down to lower temperatures. The reversal of the dominant spin sector between the region just below $T_{\mathrm c}$ and the low-temperature regime gives rise to the sign change in the quasiparticle SNC.

The negative sign of the low-temperature quasiparticle contribution originates from the opposite \(s_z\) weights of the two equal-spin sectors.
In this model, the minority-spin contribution from the $s^z_{\rm pair}=+1$ sector gives a negative contribution to \(\alpha_{xy}^{s_z}\).
At higher temperatures, this negative contribution is outweighed by the positive contribution from the majority-spin sector.
At lower temperatures, however, the majority-spin contribution with the larger excitation gap is exponentially suppressed, and the negative minority-spin contribution dominates the quasiparticle response.
This causes the sign change of the quasiparticle contribution shown in Fig.~\ref{fig4}(b).

In contrast to the quasiparticle contribution, the supercurrent contribution does not exhibit the pronounced sign reversal observed in the quasiparticle response. For the parameters considered here, $\alpha_{xy}$ and $D_{xx}^{s^z}/D_{xx}$ are both positive over the temperature range in which the supercurrent contribution is
appreciable [see Figs.~\ref{fig4}(b)--\ref{fig4}(d)].
Consequently, the supercurrent contribution $-(D_{xx}^{s^z}/D_{xx})\alpha_{xy}$ is negative.
Although $D_{xx}^{s^z}/D_{xx}$ becomes slightly negative in the low-temperature limit, the corresponding supercurrent contribution is negligibly small because $\alpha_{xy}$ is exponentially suppressed.

The temperature dependence of $D_{xx}^{s^z}/D_{xx}$ should be understood as the evolution of the spin-weighted superfluid response.
At finite temperatures, the spin sector with the larger pairing amplitude generally exhibits weaker thermal depletion and therefore makes a larger contribution to the superfluid stiffness.
Figure~\ref{fig4}(d) shows that the magnitude of $D_{xx}^{s^z}/D_{xx}$ is strongly suppressed in the zero-temperature limit, except in the fully spin-polarized case $|\eta|=1$.
At $T_{\rm eq}=0$, the superfluid density is determined by the total electron density and similarly, the spin superfluid weight is governed by the difference in the normal-state density of states between the two spin projections~\cite{tinkham_SC}.
As the normal-state Fermi surface is spin degenerate in the present model, the leading zero-temperature contributions from the two equal-spin sectors nearly cancel in the spin-weighted superfluid response.
The small negative value of $D_{xx}^{s^z}/D_{xx}$ at low $T$ may arise from a residual imbalance between the two equal-spin sectors associated with the nonunitary gap amplitudes, together with details of the band structure and particle-hole asymmetry. We note, however, that spin splitting of the normal-state bands induced by an exchange field can modify this behavior. For a wide band, the resulting difference between the two spin-resolved normal-state responses is expected to be small, whereas it may become appreciable in heavy-fermion systems with narrow bands near the chemical potential. We leave a detailed investigation of this effect for future work.

As discussed in Sec.~\ref{sec: ISNE_helicalsc}, the conserved-current formulations for the quasiparticle spin and charge currents are
essential for obtaining finite values of $\alpha_{xy}^{s^z}$ and $\alpha_{xy}$, respectively~\cite{xiao_conSC}.
The former incorporates the spin-torque-dipole contribution, whereas the latter accounts for the condensate backflow associated with charge
transfer between quasiparticles and the condensate.
As a result, the quasiparticle charge current in local equilibrium takes the form of a magnetization (circulating) current, which allows us to define the transport quasiparticle thermoelectric charge current.

\section{Conclusion}
To summarize, we theoretically investigated the intrinsic SNE in spin-triplet superconductors.
The SNE has been proposed as a probe of the helical (spin-dependent chiral) nature of the condensate in time-reversal-invariant topological superconductors~\cite{matsuhita_SNE1}.
We showed that, in spin-triplet superconductors, there are two distinct contributions to the SNE: the direct quasiparticle response is augmented by the supercurrent response to temperature gradients.
The latter is a unique feature of superconducting systems and arises from the requirement that the bulk thermoelectric charge current vanish, which induces a compensating supercurrent.

We focused on the intrinsic SNE in the clean limit, where both the quasiparticle and thermoelectric spin supercurrent responses are governed by the Berry curvature arising from spin-triplet pairing.
Consequently, the intrinsic SNE emerges and evolves below the superconducting transition temperature.
In candidate spin-triplet superconductors, the SOC-induced (normal-state) contributions are suppressed in this regime, because the superconducting gap scale is much smaller than the SOC energy scale ($T_{\mathrm c} \ll E_{\rm SOC}$).
The spin Nernst response we obtain is a hallmark of spin-triplet superconductivity. Its observation would provide clear evidence of spin-triplet Cooper pairing in a given material.

We illustrated the importance of the conserved spin current formalism by presenting an example in which this approach yields qualitatively different results compared to using the conventional spin current: a vanishing versus a finite SNE.
We further showed that both the quasiparticle and spin supercurrent responses contribute to the intrinsic SNE when the condensate is spin-polarized.
We found that the magnitudes of the two contributions are comparable in nonunitary superconductors.
This finding highlights the need to include the supercurrent response for a complete and accurate evaluation of the intrinsic SNE.

For finite impurity scattering, the extrinsic SNE can be greater than the intrinsic SNE.
This tendency is broadly consistent with previous analyses of the SNE and the ATHE in unconventional superconductors~\cite{matsushita_SNE2, matsushita_SSSE,matsushita_ATHE,wave24,sharma_ATHE,ngampruetikorn2020impurity,yip2016low,yilmaz2020spontaneous}.
Therefore, the observation of the intrinsic SNE requires sufficiently clean samples, which may be achievable in UTe$_2$~\cite{aoki_UTe2}.

For a concrete comparison with the extrinsic SNE, we refer to the previous calculation for a helical superconductor shown in Fig.~2(a) of Ref.~\cite{matsushita_SNE2}.
We focus on the case with the normal-state scattering rate $\Gamma_{\rm imp}=0.04\pi T_{\mathrm c}$ and the scattering phase shift $\delta=\pi/6$.
The phase shift $\delta=\pi/6$ corresponds to a relatively weak impurity potential compared with the Fermi energy.
Using the parameters employed in the present work, $t=1.25\,{\rm meV}$, $\mu=-2.5\,{\rm meV}$, and $T_{\mathrm c}=0.16t$, this scattering rate corresponds to an impurity concentration of approximately $1.6\%$ of the electron density.

For this parameter set, the peak value of the extrinsic contribution to the SNC is estimated to be $4.8\times 10^{-3} \left(\hbar/2e\right){\rm A}/{\rm K\,m}$.
On the other hand, the peak value of the intrinsic contribution obtained in the present work is $5.2\times 10^{-4}
\left(\hbar/2e\right){\rm A}/{\rm K\,m}$.
Therefore, for the given scattering rate and scattering phase shift, the extrinsic contribution is an order of magnitude larger than the intrinsic contribution.

\begin{figure}[t]
\includegraphics[width=70mm]{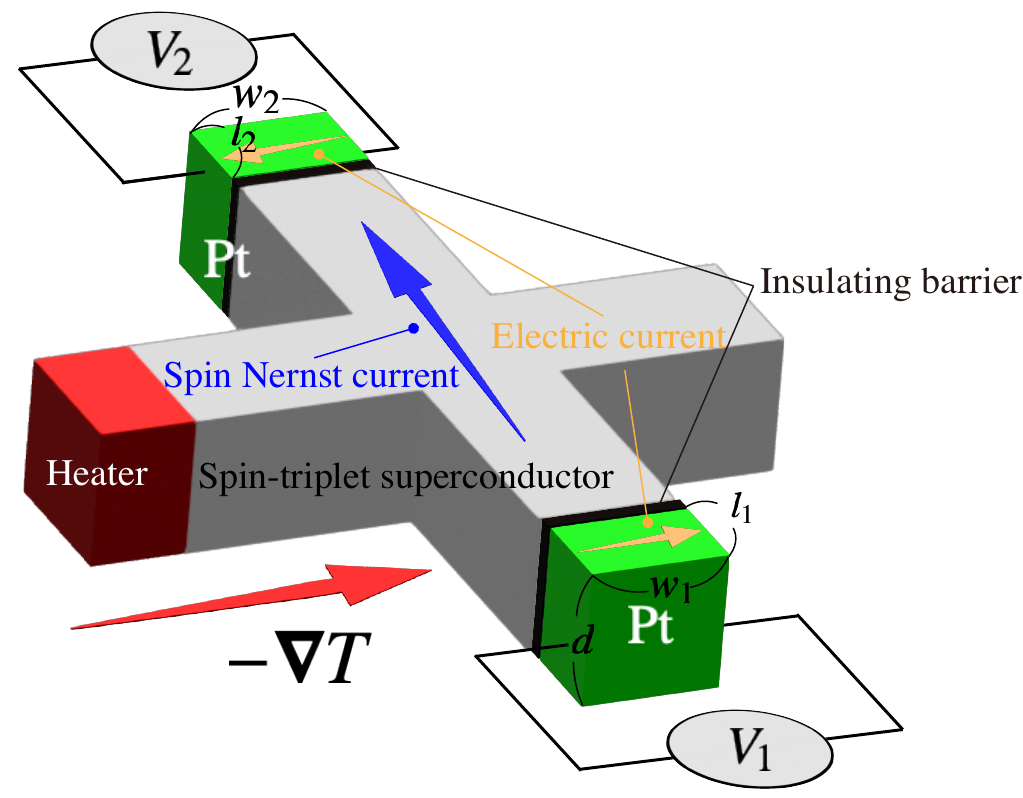}
\caption{
Schematic illustration of a possible experimental setup for detecting the SNE in spin-triplet superconductors.
A temperature gradient (red arrow) is applied to a spin-triplet superconductor (gray), generating a transverse spin Nernst current (blue arrow).
The spin current can be absorbed by a heavy metal (green) with a large spin Hall angle and converted into a charge current (orange arrows) through the ISHE.
To suppress superconducting shunting in the voltage readout, a thin insulating tunneling barrier (black) may be inserted between the superconductor and the heavy-metal detector.
}
\label{fig6}
\end{figure}

The spin current is not directly measurable, and a standard route to its detection is to convert it into an electric voltage through the inverse spin Hall effect (ISHE)~\cite{sinova_SHE}.
Figure~\ref{fig6} illustrates a possible setup for detecting the SNE in spin-triplet superconductors~\cite{bose2018,bose2019,sheng2017}.
In this configuration, the spin Nernst current generated in the spin-triplet superconductor is absorbed by an adjacent heavy metal with a large spin Hall angle, such as Pt.
The absorbed spin angular momentum is then converted into a transverse electric voltage through the ISHE, yielding a signal $V_{\rm SNE}=V_1-V_2$.
The differential measurement between $V_1$ and $V_2$ helps eliminate contributions from ordinary thermoelectric effects, such as the Seebeck effect, thereby isolating the voltage signal associated with the spin Nernst current.

In superconducting hybrid structures, however, the ISHE voltage readout can be strongly modified below $T_{\mathrm c}$ by superconducting shunting and by changes in nonequilibrium spin and charge transport. In Nb/NiFe bilayers, the ISHE voltage induced by spin pumping was found to decrease sharply below the superconducting transition temperature, a behavior attributed to the shortening of the quasiparticle charge-imbalance length in the superconducting state~\cite{jeon_spin-pumping-induced_2018}. Separately, experiments on Nb-based hybrid devices have demonstrated that superconductivity can mediate and control the transmission of pure spin currents~\cite{Jeon_prx}. These results indicate that the electrical detection of spin currents in superconducting hybrid structures is highly sensitive to the device geometry and interfacial transport processes.

In contrast to the previously studied cases, in the setup considered here, the spin current is generated in the bulk of the superconductor, while the Pt layer serves primarily as a passive spin detector. The resulting ISHE signal therefore depends critically on the boundary conditions at the superconductor/Pt interface, which determine the amount of spin current injected into the Pt layer. Whether a suitably designed insulating tunnel barrier can suppress superconducting shunting while maintaining efficient spin transfer remains an important question for future study.

As an example, we provide an order-of-magnitude estimate for uranium-based spin-triplet superconductors. We assume the following parameters: $t = 1.25$ meV (bandwidth $E_{\mathrm B}=8t=10\;\mathrm{meV}$), $\mu = -2.5$ meV, $T_{\mathrm c}=2.3$ K ($T_{\mathrm c}/t=0.16$), a temperature difference of $\Delta T = 0.2$ K applied across 1.0 mm, and platinum as a detector of the spin current~\cite{wang2014determination}.
Under these assumptions, we obtain $V_{\rm SNE}\sim 2.8$ nV for device dimensions $w_1=w_2=1.0$ mm, $l_1=l_2=1.0\times 10^2$ nm, and $d=2.0$ $\mu$m.
Assuming the same inverse-spin-Hall conversion efficiency, the extrinsic contribution estimated above gives a voltage signal of approximately $26\,{\rm nV}$.
Within this idealized estimate, both signals are potentially within experimental reach, suggesting that the intrinsic and extrinsic SNEs may be detectable in appropriately designed devices.

In addition, the spin Nernst current can be detected without converting it into an electric voltage.
Because the transverse spin current generated in the bulk relaxes near the sample edges, it induces edge spin accumulation.
Such spin accumulation can be probed by magneto-optical techniques such as the polar Kerr effect~\cite{Stamm_kerr}.
It can also be probed by local quantum-sensing techniques using NV centers~\cite{Du_spinchemical}.

Our results establish the intrinsic SNE as a sensitive probe of spin-triplet superconductivity and its underlying topological structure.
They also provide a guiding principle for identifying and characterizing unconventional superconductors through thermoelectric spin transport measurements.

\section*{Acknowledgments}
The authors thank Z.-C. Liao, C. Xiao, and Q. Niu for providing the Supplemental Material for Ref.~\cite{liao_conSC}.
T. Matsushita thanks C. Xiao, T. Kato, T. Misawa, A. Shitade, A. Daido, K. Shinada, Y. Hirobe, and J. Nasu for fruitful discussions.
This work was supported by JST CREST (Grant No.~JPMJCR19T2), JSPS KAKENHI (Grant No.~JP23K20828, No.~JP23K22492, No.~JP24KJ0130, No.~JP24H00007, No.~JP25H00599, No.~JP25H00609, No.~JP25K07227, No.~JP25K22011, No.~JP22H01181, No.~JP22H04933, No.~JP23K17353, No.~JP23K22452, No.~JP24K21530, and No.~JP26H02016), and a Grant-in-Aid for Transformative Research Areas (A) ``Correlation Design Science'' (Grant No.~JP25H01250 and No.~JP25H01249) from JSPS of Japan.
I. V. is grateful to the Aspen Center for Physics for partial support under the National Science Foundation grant PHY-2210452.

\appendix
\section{Hamiltonian and wave function of wave packet}
\label{app:semiclass}

In this Appendix, we provide a self-contained derivation of Eqs.~\eqref{eq:alpha_smu} and~\eqref{eq:alpha_mu} on the basis of the semiclassical wave packet theory.
Although these expressions were originally derived in Ref.~\cite{xiao_conSC}, we include their derivation here for completeness.

We consider the Hamiltonian acting on the wave packet~\cite{xiao_conSC,dong_semiclassical}
\begin{subequations}
\beq
\hat{\mathcal{H}}&=&\hat{\mathcal{H}}_c+\hat{\mathcal{H}}_1+\hat{\mathcal{H}}_2,\\
\hat{\mathcal{H}}_c&=&\hat{\mathcal{H}}_0(\bm r_c(t_c))+\hat{\bm \theta}\cdot \hat{\bm h}(\bm r_c(t_c)),\\
\hat{\mathcal{H}}_1&=&\frac{1}{2}\Re \left\{\hat{\theta}^\mu, r_{i}-r_{ci}\right\} \partial_{r_{ci}} h_\mu\\
\hat{\mathcal{H}}_2&=&\frac{1}{2}\Re \left\{\hat{\theta}^\mu, \frac{1}{2}(r_{i}-r_{ci})(r_{j}-r_{cj})\right\} \partial_{r_{ci}} \partial_{r_{cj}} h_\mu.
\eeq
\label{Ham_semi}
\end{subequations}
Here, $\hat{\mathcal{H}}_c$ is the Hamiltonian of a crystalline system, satisfying
\beq
\hat{\mathcal{H}}_c |\psi_{n\bm p}\rangle = \epsilon_{n\bm p} |\psi_{n\bm p}\rangle\,,
\eeq
 $|\psi_{n\bm p}\rangle = e^{i\bm p \cdot \bm r} |u_{n\bm p}\rangle$ is a Bloch state, with $|u_{n\bm p}\rangle$ its periodic part, and $\epsilon_{n\bm p}$ is the quasiparticle energy measured from the chemical potential.
The field ${\bm h}$ is slowly varying on the lattice scale, and is coupled to the conjugate physical quantity $\hat{\bm \theta}$, and $\bm r_c(t_c)$ denotes the center-of-mass coordinate of the wave packet as a function of time $t_c$.
For example, if $\hat{\bm \theta}$ represents the conventional spin current, then ${\bm h}$ corresponds to the spin-dependent vector potential. 
The explicit form of ${\bm h}$ is not necessary to calculate the equilibrium and nonequilibrium expectation values of $\hat{\bm \theta}$ because we set $\bm h \to 0$ after evaluating the transport coefficients~\cite{dong_semiclassical}.
In Eq.~\eqref{Ham_semi}, $\hat{\mathcal{H}}_1$ and $\hat{\mathcal{H}}_2$ represent the first-order and second-order terms in the expansion of the Hamiltonian with respect to ${\bm r} - {\bm r}_c$.
To evaluate $\bm J^{\bm s}$ and $\bm J^{\bm \tau}$ in the main text up to first order in the spatial gradient, we expand $\bm J^{\bm s}$ to first order and $d{\bm \tau}/dt_c$ to second order.

Let us assume that the wave packet is constructed from the $n$-th band and is slightly perturbed by the inhomogeneous $\bm h$-field.
The wave function of the wave packet is expanded to first order in $\partial_{r_{ci}} h_\mu$ and is expressed as~\cite{gao_semiclassical,xiao_conSC}
\begin{subequations}
\beq
\big|\Psi_{n\bm k} \big\rangle &=& \big|\Psi_{n\bm k}^{(0)} \big\rangle + \big|\Psi_{n\bm k}^{(1)} \big\rangle, \\
\big|\Psi_{n\bm k}^{(0)} \big\rangle &=& \int \left[d\bm p\right] C_{n\bm k}(\bm p) |\psi_{n\bm p} \rangle, \\
\big|\Psi_{n\bm k}^{(1)} \big\rangle &=& \int \left[d\bm p\right] \sum_{n_1 \ne n} C_{n_1\bm k}(\bm p) |\psi_{n_1\bm p} \rangle.
\eeq
\end{subequations}
Here, we use the shorthand notation $\left[d\bm p\right] = d\bm p / (2\pi)^D$, where $D$ is the spatial dimension.
The coefficient $C_{n\bm k}(\bm p)$ is zeroth order in $\partial_{r_{ci}} h_\mu$, while $C_{n_1\bm k}(\bm p)$ ($n_1 \ne n$) is first order.

The unperturbed wave packet exhibits a sharp peak at the center-of-mass momentum at $\bm k$ and is therefore approximated as~\cite{niu_semiclassical,xiao_berrycur}
\beq
C_{n\bm k}(\bm p) &= |C_{n\bm k}(\bm p)| e^{-i \gamma_n(\bm p)}, \
|C_{n\bm k}(\bm p)|^2 &\simeq \delta(\bm k - \bm p).
\eeq
$C_{n_1\bm k}(\bm p)$ describes the interband transition due to spatial inhomogeneity and is determined by the Schr\"{o}dinger equation
\beq
\label{eq: Schrodingereq}
i\partial_{t_c} \big|\Psi_{n\bm k}\big\rangle=\left(\hat{\mathcal{H}}_c+\hat{\mathcal{H}}_1\right)\big|\Psi_{n\bm k}\big\rangle,
\eeq
where $\hat{\mathcal{H}}_2$ is neglected because the wave function is expanded only up to the first order in the spatial gradient.
The left-hand side of Eq.~\eqref{eq: Schrodingereq} is evaluated as
\beq
\label{eq: Schrodingereq1}
\big\langle \psi_{n_1\bm p}\big|i\partial_{t_c} \Psi_{n\bm k}\big\rangle=\epsilon_{n\bm p}C_{n_1\bm k}+v^i_n(\partial_{r_{ci}} h_\mu)(\mathcal{A}_\mu^h)_{n_1n}C_{n_1\bm k},\nn\\
\eeq
for $n_1\neq n$.
Here, $(\mathcal{A}_\mu^h)_{n_1n}$ represents the interband Berry connection in the $\bm h$-space, defined as
\beq
\label{eq: Ah}
i\big\langle u_{n_1\bm p}\big| \partial_{h_\mu} u_{n\bm p}\rangle =
\begin{cases}
(\mathcal{A}_\mu^h)_{n} &n_1=n, \\
(\mathcal{A}_\mu^h)_{n_1n} &n_1\neq n.
\end{cases}
\eeq
The right-hand side of Eq.~\eqref{eq: Schrodingereq} becomes
\beq
\label{eq: Schrodingereq2}
&\big\langle \psi_{n_1\bm p}\big|\left(\hat{\mathcal{H}}_c+\hat{\mathcal{H}}_1\right)\big|\Psi_{n\bm k}\big\rangle =\epsilon_{n_1\bm p}C_{n_1\bm k}+I,
\eeq
where
\beq
\label{eq: I}
I=\int \left[d\bm p_1\right]\big\langle \psi_{n_1\bm p}\big|\hat{\mathcal{H}}_1 \big|\psi_{n\bm p_1}\big\rangle C_{n\bm k}(\bm p_1).
\eeq
By equating both sides and solving, we find
\beq
\label{eq: Cn1}
C_{n_1\bm k}=\frac{I-(\partial_{r_{ci}} h_\mu)(\mathcal{A}_\mu^h)_{n_1n}v^i_nC_{n\bm k}}{\epsilon_{n\bm p}-\epsilon_{n_1\bm p}}.
\eeq
Here, $v^i_n$ denotes the intraband velocity for the $n$-th band, defined as~\cite{xiao_berrycur}
\begin{subequations}
\beq
\label{eq: velocity}
\langle u_{n_1 \bm p} | \hat{v}^i | u_{n \bm p} \rangle &=&
\begin{cases}
v_n^i &n_1=n, \\
v_{n_1n}^i &n_1\neq n.
\end{cases},\\
v_n^i&=&\partial_{p_i} \epsilon_{n\bm p},\\
v_{n_1n}^i&=&i(\epsilon_{n_1\bm p}-\epsilon_{n\bm p})(\mathcal{A}_i^k)_{n_1n}.
\eeq
\end{subequations}
where $(\mathcal{A}_i^k)_{n_1n}$ represents the interband Berry connection in the momentum space, defined as
\beq
\label{eq: Ak}
i\big\langle u_{n_1\bm k}\big| \partial_{k_i} u_{n\bm k}\rangle =
\begin{cases}
(\mathcal{A}_i^k)_{n} &n_1=n, \\
(\mathcal{A}_i^k)_{n_1n} &n_1\neq n.
\end{cases}
\eeq
An explicit evaluation of $I$ gives~\cite{gao_semiclassical,xiao_conSC}
\beq
\label{eq: I1}
I&=&-i(\partial_{r_{ci}}h_\mu)\left(\epsilon_{n\bm p}-\epsilon_{n_1\bm p}\right)\nn\\
&\times&\left[\left(i\partial_{r_{ci}}+\left(\mathcal{A}_i^k\right)_n -r_{ci}\right)C_{n\bm k}\right]\left(\mathcal{A}_\mu^h\right)_{n_1n}\nn\\
&+&(\partial_{r_{ci}}h_\mu)\left(d_i^{\theta^\mu}\right)_{n_1n}C_{n\bm k}+(\partial_{r_{ci}}h_\mu)\left(\mathcal{A}_\mu^h\right)_{n_1n}v^i_nC_{n\bm k},\nn\\
\eeq
where $\left(d_i^{\theta^\mu}\right)_{n_1n}$ represents the interband dipole moment of $\hat{\theta}^\mu$ and is defined as~\cite{dong_semiclassical}
\begin{subequations}
\beq
\label{eq: dthetan}
\left(d_i^{\theta^\mu}\right)_{n}&=&\Re \sum_{m(\neq n)} (\mathcal{A}_i^k)_{nm}(\theta^\mu)_{mn},\\
\label{eq: dthetan1n}
\left(d_i^{\theta^\mu}\right)_{n_1n}&=&\frac{i}{2}\partial_{p_i}(\theta^\mu)_{n_1n}\nn\\
&+&\frac{1}{2}\sum_{n_2}\left[(\mathcal{A}_i^k)_{n_1n_2}(\theta^\mu)_{n_2n}+(\theta^\mu)_{n_1n_2}(\mathcal{A}_i^k)_{n_2n}\right]\nn\\
&-&(\theta^\mu)_{n_1n}(\mathcal{A}_i^k)_{n}-(\mathcal{A}_\mu^h)_{n}v_n^i,\;\;(n_1\neq n)
\eeq
\end{subequations}
where 
\beq
\langle u_{m\bm p}|\hat{\theta}^\mu| u_{n\bm p}\rangle=
\begin{cases}
(\theta^\mu)_{n} & n=m,\\
(\theta^\mu)_{mn} & n\neq m.
\end{cases}
\eeq
To derive Eq.~\eqref{eq: dthetan}, we use the following relationship:~\cite{xiao_berrycur}
\beq
\label{eq: position}
\langle \psi_{n_1\bm p_1}|r_i|\psi_{n\bm p}\rangle =\left[-i {\partial}_{p_i}\delta_{n_1n} + (\mathcal{A}_i^k)_{n_1n}\right] \delta(\bm p-\bm p_1).\nn\\
\eeq
Substituting Eq.~\eqref{eq: I1} into Eq.~\eqref{eq: Schrodingereq2}, we obtain
\beq
\label{eq: Cn1_1}
C_{n_1\bm k}&=&-i(\partial_{r_{ci}}h_\mu)\left[\left(i\partial_{r_{ci}}+\left(\mathcal{A}_i^k\right)_n -r_{ci}\right)C_{n\bm k}\right]\left(\mathcal{A}_\mu^h\right)_{n_1n}\nn\\
&+&\frac{(\partial_{r_{ci}}h_\mu)\left(d_i^{\theta^\mu}\right)_{n_1n}}{\epsilon_{n\bm p}-\epsilon_{n_1\bm p}}C_{n\bm k}.
\eeq

With the wave function of the wave packet, the center of mass coordinates are evaluated as
\beq
r_{ci}&=&\langle \Psi_{n\bm k}|r_i|\Psi_{n\bm k}\rangle \nn\\
&=&\int [d\bm p_1][d\bm p_2]\sum_{n_1,n_2}C_{n_1\bm k}^\ast(\bm p_1) C_{n_2\bm k}(\bm p_2) \langle \psi_{n\bm p_1}|r_i|\psi_{n\bm p_2}\rangle \nn\\
&=&\int [d\bm p_1][d\bm p_2]\sum_{n_1,n_2}C_{n_1\bm k}^\ast(\bm p_1) C_{n_2\bm k}(\bm p_2)\nn\\
&\times&\left[\left(-i\partial_{p_{2i}} \delta_{n_1n_2}+(\mathcal{A}_i^k)_{n_1n_2}\right)\delta (\bm p_2-\bm p_1) \right]\nn\\
&\simeq& \partial_{k_i}\gamma_n+(\mathcal{A}_i^k)_{n}+(\delta \mathcal{A}_i^k)_{n},
\eeq
where the positional shift (i.e., the correction to the Berry connection in the momentum space) is defined as~\cite{gao_positionalshift,gao_semiclassical}
\beq
\label{eq: positionalshift_k}
(\delta \mathcal{A}_i^k)_{n}=2\Re \int [d\bm p] \sum_{n_1\neq n} C_{n\bm k}^\ast(\bm p) C_{n_1\bm k}(\bm p)(\mathcal{A}_i^k)_{nn_1}.\nn\\
\eeq
With Eq.~\eqref{eq: Cn1_1}, the positional shift is written as
\beq
\label{eq: positionalshift_k1}
(\delta \mathcal{A}_i^k)_{n}&=&2\Re \sum_{n_1(\neq n)}\left[\frac{\left(\mathcal{A}_i^k\right)_{nn_1}\left(d_j^{\theta^\mu}\right)_{n_1n}}{\epsilon_{n\bm k}-\epsilon_{n_1\bm k}}\right](\partial_{r_{cj}} h_\mu)\nn\\
&-&\partial_{k_j}(g_{i\mu}^{kh})_n (\partial_{r_{cj}} h_\mu),
\eeq
where $(g_{ij}^{kh})_n$ is the intraband quantum metric in $\bm k$-$\bm h$ space for the $n$-th band, defined as
\beq
\label{eq: metric}
(g_{i\mu}^{kh})_n&=&\Re \sum_{n_1(\neq n)} \left(\mathcal{A}_i^k\right)_{nn_1}\left(\mathcal{A}_\mu^h\right)_{n_1n}\nn\\
&=& \Re \langle \partial_{k_i} u_{n\bm k}| \partial_{h_\mu} u_{n\bm k}\rangle - \left(\mathcal{A}_i^k\right)_{n}\left(\mathcal{A}_\mu^h\right)_{n}.
\eeq

\section{Effective Lagrangian of wave packet}
Our goal is to describe the dynamics of the wave packet by the equations of motion, involving the time-dependent momentum and center-of-mass coordinate~\cite{niu_semiclassical,xiao_berrycur}.
This allows for capturing the essential dynamics using a small number of parameters.
The equations of motion need to be derived from the effective Lagrangian via the principle of least action. 
Hence, we begin by writing this effective Lagrangian of the wave packet as
\beq
\mathcal{L}=\langle \Psi_{n\bm k}|\left(i\partial_{t_c}-\hat{\mathcal{H}}\right)|\Psi_{n\bm k}\rangle.
\eeq
We evaluate this expression term by term.
The time derivative term reads
\beq
\label{eq: part}
\langle \Psi_{n\bm k}|i\partial_{t_c}|\Psi_{n\bm k}\rangle
&=&\int [d\bm p][d\bm p_1]\sum_{n_1,n_2}\big\langle \psi_{n\bm p}\big|C_{n_1\bm k}^\ast (\bm p)\nn\\
&\times&\left[\left(i\partial_{t_c} C_{n_2\bm k}\right)+iC_{n_2\bm k}\left(\dot{r}_{ci}\partial_{r_{ci}}+\partial_{t_c}\right)\right]\big|\psi_{n_2\bm p}\big\rangle.\nn\\
\eeq
After evaluating the integrals, we find
\beq
\langle \Psi_{n\bm k}|i\partial_{t_c}|\Psi_{n\bm k}\rangle
&=&\partial_{t_c} \gamma_n+\dot{r}_{ci} \left(\tilde{\mathcal{A}}_i^r\right)_n+\left(\tilde{\mathcal{A}}^t\right)_n\nn\\
&=&\frac{d\gamma_n}{dt_c} - \dot{k}_i r_{ci}+\dot{k}_i \left(\tilde{\mathcal{A}}_i^k\right)_n\nn\\
&+&\dot{r}_{ci} \left(\tilde{\mathcal{A}}_i^r\right)_n+\left(\tilde{\mathcal{A}}^t\right)_n,\nn\\
&=&\frac{d\gamma_n}{dt_c} - \dot{k}_i r_{ci}+\dot{k}_i \left(\tilde{\mathcal{A}}_i^k\right)_n\nn\\
&+&\left(\dot{r}_{ci}\partial_{r_i}h_\mu+\partial_{t_c}h_\mu\right) \left(\tilde{\mathcal{A}}_\mu^h\right)_n.
\eeq
Here, we introduced the shorthand notation $\dot{X}=dX/dt_c$ for any time-dependent quantity $X=X(t_c)$.
The modified Berry connections are given by  
\begin{subequations}
\beq
(\tilde{\mathcal{A}}_i^\lambda)_n&=&(\mathcal{A}_i^\lambda)_n+(\delta \mathcal{A}_i^\lambda)_n,\\
(\tilde{\mathcal{A}}^t)_n&=&(\mathcal{A}^t)_n+(\delta \mathcal{A}^t)_n,
\eeq
\end{subequations}
where
\begin{subequations}
\beq
\label{eq: Ar}
i\big\langle u_{n_1\bm p}\big| \partial_{\lambda_{i}} u_{n\bm p}\rangle &=&
\begin{cases}
(\mathcal{A}_i^\lambda)_{n} &n_1=n, \\
(\mathcal{A}_i^\lambda)_{n_1n} &n_1\neq n.
\end{cases},\\
\label{eq: At}
i\big\langle u_{n_1\bm p}\big| \partial_{t_{c}} u_{n\bm p}\rangle &=&
\begin{cases}
(\mathcal{A}^t)_{n} &n_1=n, \\
(\mathcal{A}^t)_{n_1n} &n_1\neq n.
\end{cases},
\eeq
\end{subequations}
\begin{subequations}
\beq
\label{eq: positionalshift_r}
(\delta \mathcal{A}_i^\lambda)_{n}&=&2\Re \int [d\bm p] \sum_{n_1\neq n} C_{n\bm k}^\ast C_{n_1\bm k}(\mathcal{A}_i^\lambda)_{nn_1},\\
\label{eq: positionalshift_t}
(\delta \mathcal{A}^t)_{n}&=&2\Re \int [d\bm p] \sum_{n_1\neq n} C_{n\bm k}^\ast C_{n_1\bm k}(\mathcal{A}^t)_{nn_1},
\eeq
\end{subequations}
with $\lambda=r,k,h$.
Note that total time-derivative terms do not affect the equations of motion and can be omitted from the Lagrangian.
Hence, the effective Lagrangian is simplified to
\beq
\label{eq: lagrangian}
\mathcal{L}&=&k_i \dot{r}_{ci}+\dot{k}_i \left(\tilde{\mathcal{A}}_i^k\right)_n+\dot{r}_{ci} \left(\tilde{\mathcal{A}}_i^r\right)_n+\left(\tilde{\mathcal{A}}^t\right)_n-\tilde{\epsilon}_{n\bm k}\nn\\
&=&k_i \dot{r}_{ci}+\dot{k}_i \left(\tilde{\mathcal{A}}_i^k\right)_n+\left(\dot{r}_{ci}\partial_i h_\mu +\partial_{t_c}h_\mu\right) \left(\tilde{\mathcal{A}}_\mu^h\right)_n-\tilde{\epsilon}_{n\bm k},\nn\\
\eeq
where $\tilde{\epsilon}_{n\bm k}=\langle \Psi_{n\bm k}|\hat{\mathcal{H}}|\Psi_{n\bm k}\rangle$ denotes the energy of the wave packet.

\section{Equation of motion of wave packet}
From the effective action of the wave packet
\beq
\mathcal{S}=\int dt_c \mathcal{L},
\eeq
we obtain the Euler-Lagrange equations
\begin{subequations}
\beq
\frac{d}{dt_c}\frac{\partial \mathcal{L}}{\partial \dot{r_{ci}}}&=&\frac{\partial \mathcal{L}}{\partial r_{ci}},\\
\frac{d}{dt_c}\frac{\partial \mathcal{L}}{\partial \dot{k_i}}&=&\frac{\partial \mathcal{L}}{\partial k_i}.
\eeq
\end{subequations}
Differentiating the effective action (Eq.~\eqref{eq: lagrangian}), we arrive at the equations of motion for the wave packet~\cite{xiao_berrycur,gao_semiclassical}
\begin{subequations}
\beq
\dot{r}_{ci}&=&\partial_{k_i}\tilde{\epsilon}_{n\bm k}-\left[\left(\tilde{\mathcal{B}}_{ij}^{kk}\right)_n\dot{k}_j+\left(\tilde{\mathcal{B}}_{ij}^{kr}\right)_n\dot{r}_{cj}+\left(\tilde{\mathcal{E}}_{i}^{k}\right)_n\right],\\
\dot{k}_i&=&-\partial_{r_{ci}}\tilde{\epsilon}_{n\bm k}+\left[\left(\tilde{\mathcal{B}}_{ij}^{rk}\right)_n\dot{k}_j+\left(\tilde{\mathcal{B}}_{ij}^{rr}\right)_n\dot{r}_j+\left(\tilde{\mathcal{E}}_{i}^{r}\right)_n\right].
\eeq
\end{subequations}
Here, the modified Berry curvatures for the $n$-th band are defined as
\begin{subequations}
\beq
\left(\tilde{\mathcal{B}}_{ij}^{\lambda \rho}\right)_n&=&\left(\mathcal{B}_{ij}^{\lambda \rho}\right)_n+\left(\delta \mathcal{B}_{ij}^{\lambda \rho}\right)_n,\\
\left(\tilde{\mathcal{E}}_{i}^{\lambda}\right)_n&=&\left(\mathcal{E}_{i}^{\lambda}\right)_n+\left(\delta \mathcal{E}_{i}^{\lambda}\right)_n,
\eeq
\end{subequations}
where 
\begin{subequations}
\beq
\left(\mathcal{B}_{ij}^{\lambda \rho}\right)_n&=&\partial_{\lambda_i} \left(\mathcal{A}_{j}^{\rho}\right)_n-\partial_{\rho_j} \left(\mathcal{A}_{i}^{\lambda}\right)_n,\\
\left(\delta \mathcal{B}_{ij}^{\lambda \rho}\right)_n&=&\partial_{\lambda_i} \left(\delta \mathcal{A}_{j}^{\rho}\right)_n-\partial_{\rho_j} \left(\delta \mathcal{A}_{i}^{\lambda}\right)_n,\\
\left(\mathcal{E}_{i}^{\lambda}\right)_n&=&\partial_{\lambda_{i}} \left(\mathcal{A}^{t}\right)_n-\partial_{t_{c}} \left(\mathcal{A}_i^{\lambda}\right)_n,\\
\left(\delta \mathcal{E}_{i}^{\lambda}\right)_n&=&\partial_{\lambda_{i}} \left(\delta \mathcal{A}^{t}\right)_n-\partial_{t_{c}} \left(\delta \mathcal{A}_i^{\lambda}\right)_n,
\eeq
with $\lambda,\rho=r,k,h$.
\end{subequations}

\section{Energy of wave packet}
To analyze the dynamics of the wave packet, we evaluate its energy.
Expanding the energy of the wave packet up to the first order in $\bm h$ and the second order in the spatial gradient, we obtain
\beq
\label{eq: wpenergy}
\tilde{\epsilon}_{n\bm k}=\epsilon_{n\bm k}+\big\langle \Psi_{n\bm k}^{(0)}\big|\hat{\mathcal{H}}_1\big| \Psi_{n\bm k}^{(0)} \big\rangle+\big\langle \Psi_{n\bm k}^{(0)}\big|\hat{\mathcal{H}}_2\big| \Psi_{n\bm k}^{(0)} \big\rangle.
\eeq
With Eq.~\eqref{eq: position}, the second term in Eq.~\eqref{eq: wpenergy} is evaluated as
\beq
\label{eq: dipole_energy}
\big\langle \Psi_{n\bm k}^{(0)}\big|\hat{\mathcal{H}}_1\big| \Psi_{n\bm k}^{(0)} \big\rangle
\simeq \left(d_i^{\theta^\mu}\right)_n(\partial_{r_{ci}}h_\mu).
\eeq
This term represents the correction to the energy of the wave packet due to the coupling between the dipole moment of $\hat{\bm \theta}$ and the inhomogeneous auxiliary field $\bm h$~\cite{dong_semiclassical}.
The third term in Eq.~\eqref{eq: wpenergy} is given by
\beq
\label{eq: quadrupole_energy}
\big\langle \Psi_{n\bm k}^{(0)}\big|\hat{\mathcal{H}}_2\big| \Psi_{n\bm k}^{(0)} \big\rangle
= \left(q_{ij}^{\theta^\mu}\right)_n(\partial_{r_{ci}}\partial_{r_{cj}}h_\mu),
\eeq
where $\left(q_{ij}^{\theta^\mu}\right)_n$ represents the quadrupole moment of $\hat{\bm \theta}$, defined as~\cite{xiao_conSC}
\beq
\label{eq: quadrupole}
\left(q_{ij}^{\theta^\mu}\right)_n&=&\frac{1}{2}\Re \big\langle \Psi_{n\bm k}^{(0)}\big|(r_i-r_{ci})(r_j-r_{cj})\hat{\theta}^\mu\big|\Psi^{(0)}_{n\bm k}\big\rangle \nn\\
&=&\frac{1}{2}\Re \langle \Psi^{(0)}_{n\bm k}|r_ir_j\hat{\theta}^\mu|\Psi^{(0)}_{n\bm k}\rangle-\frac{1}{2}r_{ci}\left(d_j^{\theta^\mu}\right)_n \nn\\
&-& \frac{1}{2}r_{cj}\left(d_i^{\theta^\mu}\right)_n+\frac{1}{2}r_{ci}r_{cj}(\theta^\mu)_n.
\eeq
Using the identity~\cite{lapa_semiclassical}
\beq
&\langle \psi_{n_1\bm p_1}|r_ir_j|\psi_{n\bm p}\rangle =\partial_{p_{1i}}\partial_{p_j}\delta_{n_1n}-i(\mathcal{A}_i^k)_{n_1n}\left(\partial_{p_j} \delta(\bm p_1-\bm p)\right)\nn\\
&-i(\mathcal{A}_j^k)_{n_1n}\left(\partial_{p_{1i}} \delta(\bm p_1-\bm p)\right)-\delta(\bm p_1-\bm p)\langle u_{n_1\bm p_1}|\partial_{p_{1i}}\partial_{p_j} u_{n \bm p}\rangle,\nn\\
\eeq
we evaluate the first term in Eq.~\eqref{eq: quadrupole} as
\beq
\label{eq: rirj}
&\Re& \langle \Psi^{(0)}_{n\bm k}|r_ir_j\hat{\theta}^\mu|\Psi^{(0)}_{n\bm k}\rangle\nn\\
&=&(\partial_{k_i}\gamma_n)(\partial_{k_j}\gamma_n)(\theta^\mu)_n-\partial_{k_i}\partial_{k_j}(\theta^\mu)_n\nn\\
&+&\sum_{n_1}\big[(\partial_{k_j}\gamma_n)(\mathcal{A}_i^k)_{nn_1}(\theta^\mu)_{n_1n}+(\partial_{k_i}\gamma_n)(\mathcal{A}_j^k)_{nn_1}(\theta^\mu)_{n_1n}\big]\nn\\
&+&\sum_{n_1}\big[i(\mathcal{A}_i^k)_{nn_1}\partial_{k_j}(\theta^\mu)_{n_1n}+i(\mathcal{A}_j^k)_{nn_1}\partial_{k_i}(\theta^\mu)_{n_1n}\nn\\
&-&\langle u_{n_1\bm k}|\partial_{k_{i}}\partial_{k_j} u_{n \bm k}\rangle (\theta^\mu)_{n_1n}\big].
\eeq
Using
\beq
\partial_{k_i}\partial_{k_j}(\theta^\mu)_n &=&\langle \partial_{k_i}\partial_{k_j}u_{n\bm k}|\hat{\theta}^\mu|u_{n\bm k}\rangle+\langle \partial_{k_j}u_{n\bm k}|\left(\partial_{k_i}\hat{\theta}^\mu|u_{n\bm k}\rangle \right)\nn\\
&+&\langle \partial_{k_i}u_{n\bm k}|\left(\partial_{k_j}\hat{\theta}^\mu|u_{n\bm k}\rangle \right)\nn\\
&+&\langle u_{n\bm k}|\left(\partial_{k_i}\partial_{k_j}\hat{\theta}^\mu|u_{n\bm k}\rangle \right),
\eeq
the last three terms in Eq.~\eqref{eq: rirj} are rewritten as
\beq
\label{eq: rirj1}
&&\sum_{n_1}\bigg[i(\mathcal{A}_i^k)_{nn_1}\partial_{k_j}(\theta^\mu)_{n_1n}+i(\mathcal{A}_j^k)_{nn_1}\partial_{k_i}(\theta^\mu)_{n_1n}\nn\\
&-&\langle u_{n_1\bm k}|\partial_{k_{i}}\partial_{k_j} u_{n \bm k}\rangle (\theta^\mu)_{n_1n}\bigg]\nn\\
&=&\frac{1}{2}\Re \bigg[\langle \partial_{k_j}u_{n\bm k}|\left(\partial_{k_i}\hat{\theta}^\mu|u_{n\bm k}\rangle \right)+\langle \partial_{k_i}u_{n\bm k}|\left(\partial_{k_j}\hat{\theta}^\mu|u_{n\bm k}\rangle \right)\nn\\
&+&\frac{1}{2}\partial_{k_i}\partial_{k_j}(\theta^\mu)_n\bigg].
\eeq
Substituting Eq.~\eqref{eq: rirj1} into Eq.~\eqref{eq: quadrupole}, we obtain $\left(q_{ij}^{\theta^\mu}\right)_n$ as
\begin{widetext}
\beq
\label{eq: quadrupole1}
2\left(q_{ij}^{\theta^\mu}\right)_n&=&(\mathcal{A}_i^k)_{n}(\mathcal{A}_j^k)_{n}(\theta^\mu)_n-\frac{1}{2}\partial_{k_i}\partial_{k_j}(\theta^\mu)_n-(\mathcal{A}_i^k)_{n}\Re \sum_{n_1}(\mathcal{A}_j^k)_{nn_1}(\theta^\mu)_{n_1n}-(\mathcal{A}_j^k)_{n}\Re \sum_{n_1}(\mathcal{A}_i^k)_{nn_1}(\theta^\mu)_{n_1n}\nn\\
&+&\frac{1}{2}\Re \left[
\langle \partial_{k_j}u_{n\bm k}|\left(\partial_{k_i}\hat{\theta}^\mu\right)|u_{n\bm k}\rangle +\langle \partial_{k_i}u_{n\bm k}|\left(\partial_{k_j}\hat{\theta}^\mu\right)|u_{n\bm k}\rangle \right]\nn\\
&+&\frac{1}{2}\Re \sum_{n_1,n_2}\left[(\mathcal{A}_i^k)_{nn_1}(\theta^\mu)_{n_1n_2}(\mathcal{A}_j^k)_{n_2n}+(\mathcal{A}_j^k)_{nn_1}(\theta^\mu)_{n_1n_2}(\mathcal{A}_i^k)_{n_2n}\right].
\eeq
Reorganizing the last two terms in Eq.~\eqref{eq: quadrupole1}
\beq
&&\frac{1}{2}\Re \sum_{n_1,n_2}\left[(\mathcal{A}_i^k)_{nn_1}(\theta^\mu)_{n_1n_2}(\mathcal{A}_j^k)_{n_2n}+(\mathcal{A}_j^k)_{nn_1}(\theta^\mu)_{n_1n_2}(\mathcal{A}_i^k)_{n_2n}\right]\nn\\
&=&\Re \sum_{n_1(\neq n),n_2 (\neq n)}(\mathcal{A}_i^k)_{nn_1}(\theta^\mu)_{n_1n_2}(\mathcal{A}_j^k)_{n_2n}-(\mathcal{A}_i^k)_{n}(\mathcal{A}_j^k)_{n}(\theta^\mu)_n+\Re \sum_{n_1}\left[(\mathcal{A}_i^k)_{n}(\mathcal{A}_j^k)_{nn_1}(\theta^\mu)_{n_1n}+(\mathcal{A}_i^k)_{nn_1}(\theta^\mu)_{n_1n}(\mathcal{A}_j^k)_{n}\right],\nn\\
\eeq
we finally obtain~\cite{xiao_conSC}
\beq
\label{eq: quadrupole2}
\left(q_{ij}^{\theta^\mu}\right)_n&=&-\frac{1}{4}\partial_{k_i}\partial_{k_j}(\theta^\mu)_n
+\frac{1}{2}\Re \sum_{n_1\neq n,n_2 \neq n}(\mathcal{A}_i^k)_{nn_1}(\theta^\mu)_{n_1n_2}(\mathcal{A}_j^k)_{n_2n}+\frac{1}{4}\Re \left[
\langle \partial_{k_j}u_{n\bm k}|\left(\partial_{k_i}\hat{\theta}^\mu\right)|u_{n\bm k}\rangle +\langle \partial_{k_i}u_{n\bm k}|\left(\partial_{k_j}\hat{\theta}^\mu\right)|u_{n\bm k}\rangle \right].\nn\\
\eeq
\end{widetext}

\section{Quadrupole moment of $\hat{\theta}^\mu=d\hat{\Theta}^\mu/dt_c$}
We now derive the expression for the quadrupole moment of $\hat{\theta}^\mu=d\hat{\Theta}^\mu/dt_c$.
Because $(d\hat{\Theta}^\mu/dt_c)_n=0$, Eq.~\eqref{eq: quadrupole2} is simplified to
\beq
\label{eq: quadrupole3}
&&4\left(q_{ij}^{\theta^\mu}\right)_n=2\Re \sum_{n_1(\neq n),n_2 (\neq n)}(\mathcal{A}_i^k)_{nn_1}(\theta^\mu)_{n_1n_2}(\mathcal{A}_j^k)_{n_2n}\nn\\
&+&\Re \left[
\langle \partial_{k_j}u_{n\bm k}|\left(\partial_{k_i}\hat{\theta}^\mu\right)|u_{n\bm k}\rangle +\langle \partial_{k_i}u_{n\bm k}|\left(\partial_{k_j}\hat{\theta}^\mu\right)|u_{n\bm k}\rangle \right].\nn\\
\eeq
Using the identity
\beq
\partial_{k_i}\left(\hat{\theta}^\mu\right)_{n_1n}&=&
\langle u_{n_1\bm k}|\left(\partial_{k_i}\hat{\theta}^\mu\right)|u_{n\bm k}\rangle\nn\\
&+&i\sum_{n_2}\left[(\mathcal{A}_i^k)_{n_1n_2}\left(\hat{\theta}^\mu\right)_{n_2n}-\left(\hat{\theta}^\mu\right)_{n_1n_2}(\mathcal{A}_i^k)_{n_2n}\right],\nn\\
\eeq
we reorganize Eq.~\eqref{eq: quadrupole3} as
\beq
\label{eq: quadrupole4}
4\left(q_{ij}^{\theta^\mu}\right)_n
&=&
\Re \sum_{n_1(\neq n)}(\mathcal{A}_i^k)_{nn_1}\left[\partial_{k_j}(\Theta^\mu)_{n_1n}\right](\epsilon_{n\bm k}-\epsilon_{n_1\bm k})\nn\\
&+&\Re \sum_{n_1(\neq n)}(\mathcal{A}_i^k)_{nn_1}(\Theta^\mu)_{n_1n}(v_{n}^j-v_{n_1}^j)\nn\\
&+&\Re \sum_{n_1(\neq n),n_2(\neq n,n_1)}(\mathcal{A}_i^k)_{nn_1}(\mathcal{A}_j^k)_{n_1n_2}(\theta^\mu)_{n_2n}\nn\\
&+&\Re \sum_{n_1(\neq n)}(\mathcal{A}_i^k)_{nn_1}(\theta^\mu)_{n_1n}(\mathcal{A}_j^k)_{n_1}\nn\\
&-&\Re \sum_{n_1(\neq n)}(\mathcal{A}_i^k)_{nn_1}(\theta^\mu)_{n_1n}(\mathcal{A}_j^k)_{n}\nn\\
&+&(i\leftrightarrow j).
\eeq
Upon substituting the identity
\beq
\label{eq: delTheta}
\partial_{k_i}\Theta^\mu_{n_1n}=i\sum_{n_2}\left[(\mathcal{A}_i^k)_{n_1n_2}\Theta^\mu_{n_2n}-\Theta^\mu_{n_1n_2}(\mathcal{A}_i^k)_{n_2n}\right],
\eeq
into Eq.~\eqref{eq: quadrupole4} and performing some algebra, we obtain
\beq
\label{eq: quadrupole5}
\left(q_{ij}^{\theta^\mu}\right)_n
=-\frac{1}{2}\left(\mathcal{B}_{ij}^{kk\Theta^\mu}\right)_n+\left(d_i^{\Theta^\mu}\right)_{n}v_n^j-\left(m_{ij}^{\Theta^\mu}\right)_n,
\eeq
where the orbital angular momentum of $\hat{\Theta}_l$ is defined as~\cite{xiao_conSC}
\beq
\left(m_{ij}^{\Theta^\mu}\right)_n&=&\frac{1}{2}\Re \sum_{n_1(\neq n)}
\bigg[  
(\mathcal{A}_i^k)_{nn_1}\left(J_j^{\Theta^\mu}\right)_{n_1n}\nn\\
&-&(\mathcal{A}_j^k)_{nn_1}\left(J_i^{\Theta^\mu}\right)_{n_1n}
\bigg]\nn\\
&+&\frac{1}{2}\left[
\left(d_i^{\Theta^\mu}\right)_{n}v_n^j-\left(d_j^{\Theta^\mu}\right)_{n}v_n^i
\right].
\eeq

\section{Positional shift when $\hat{\theta}^\mu=d\hat{\Theta}^\mu/dt_c$}
We evaluate the positional shift [Eq.~\eqref{eq: positionalshift_k1}] for a case of $\hat{\theta}^\mu=d\hat{\Theta}^\mu/dt_c$, where $\hat{\Theta}^\mu$ is a physical quantity.
Because $d\hat{\Theta}^\mu/dt_c=(1/i)[\hat{\Theta}^\mu,\hat{\mathcal{H}}]$, the interband $\bm h$-space Berry connection is equivalent to the interband matrix element of $\hat{\Theta}^\mu$
\beq
\label{eq: Ahtau}
(\mathcal{A}_\mu^h)_{n_1n}=\Theta^\mu_{n_1n}, 
\eeq
where $\Theta^\mu_{n_1n}=\langle u_{n_1 \bm k} | \hat{\Theta}^\mu | u_{n \bm k} \rangle$ ($n_1\neq n$).
Hence, the $\bm k-\bm h$ space quantum metric is expressed as
\beq
(g_{i\mu}^{kh})_n=\Im \sum_{n_1(\neq n)} \frac{v^i_{nn_1}\Theta^\mu_{n_1n}}{\epsilon_{n\bm k}-\epsilon_{n_1\bm k}}.
\eeq
From Eq.~\eqref{eq: dthetan}, the $\bm k-\bm h$ space quantum metric is equal to the dipole moment of $\hat{\Theta}^\mu$
\beq
\label{eq: metric1}
(g_{i\mu}^{kh})_n=\left(d_i^{\Theta^\mu}\right)_{n}.
\eeq
Using Eq.~\eqref{eq: dthetan1n} and $\hat{\theta}^\mu = d\hat{\Theta}^\mu / dt_c$, we evaluate $\left(d_i^{\theta^\mu} \right)_{n_1 n}$ as
\beq
\label{eq: dthetatau}
\left(d_i^{\theta^\mu}\right)_{n_1n}&=&-\frac{1}{2}\Theta^\mu_{n_1n}(v^i_{n}+v^i_{n_1})+\left[\partial_{k_i}\Theta^\mu_{n_1n}\right](\epsilon_{n\bm k}-\epsilon_{n_1\bm k})\nn\\
&-&\frac{i}{2}\left[(\mathcal{A}_i^k)_{n_1}+(\mathcal{A}_i^k)_{n}\right]\Theta^\mu_{n_1n}(\epsilon_{n\bm k}-\epsilon_{n_1\bm k})\nn\\
&-&\frac{i}{2}\sum_{n_2(\neq n_1)}(\mathcal{A}_i^k)_{n_1n_2}\Theta^\mu_{n_2n}(\epsilon_{n\bm k}-\epsilon_{n_2\bm k})\nn\\
&-&\frac{i}{2}\sum_{n_2(\neq n_1)}\Theta^\mu_{n_1n_2}(\mathcal{A}_i^k)_{n_2n}(\epsilon_{n_2\bm k}-\epsilon_{n_1\bm k}).
\eeq
Using Eq.~\eqref{eq: delTheta}, we can rewrite Eq.~\eqref{eq: dthetatau} as
\beq
\label{eq: dthetatau1}
\left(d_i^{\theta^\mu}\right)_{n_1n}=-\left(J_i^{\Theta^\mu}\right)_{n_1n}+i\sum_{n_2(\neq n)}\Theta^\mu_{n_1n_2}(\mathcal{A}_i^k)_{n_2n}(\epsilon_{n\bm k}-\epsilon_{n_1\bm k}),\nn\\
\eeq
where $\left(J_i^{\Theta^\mu}\right)_{n_1n}=\langle u_{n_1 \bm k} | \hat{J}_i^{\Theta^\mu} | u_{n \bm k} \rangle$ is the matrix element of the operator $\hat{J}_i^{\Theta^\mu}=\{\hat{v}_i,\hat{\Theta}^\mu\}/2$.
Substituting Eqs.~\eqref{eq: metric1} and \eqref{eq: dthetatau1} into Eq.~\eqref{eq: positionalshift_k1}, we obtain
\beq
\label{eq: positionalshift_k2}
\frac{\partial (\delta \mathcal{A}_i^k)_{n}}{\partial (\partial_{r_{cj}}h_\mu)}&=&\left(\mathcal{B}_{ij}^{kk\Theta^\mu}\right)_n-\partial_{k_j}\left(d_i^{\Theta^\mu}\right)_{n}\nn\\
&+&2\Im \sum_{n_1(\neq n), n_2(\neq n)}(\mathcal{A}_i^k)_{nn_1}\Theta^\mu_{n_1n_2}(\mathcal{A}_j^k)_{n_2n},\nn\\
\eeq
where the $\Theta$-Berry curvature is defined as
\beq
\left(\mathcal{B}_{ij}^{kk\Theta^\mu}\right)_n=-2\Im \sum_{n_1(\neq n)}\frac{v^i_{nn_1}\left(J_j^{\Theta^\mu}\right)_{n_1n}}{\left(\epsilon_{n\bm k}-\epsilon_{n_1\bm k}\right)^2}.
\eeq
We further reorganize the last two terms in Eq~\eqref{eq: positionalshift_k2}.
We rewrite the second term as
\beq
\partial_{k_j}\left(d_i^{\Theta^\mu}\right)_{n}=\left[\partial_{k_j}\left(d_i^{\Theta^\mu}\right)_{n}-\partial_{k_i}\left(d_j^{\Theta^\mu}\right)_{n}\right]+\partial_{k_i}\left(d_j^{\Theta^\mu}\right)_{n}.\nn\\
\eeq
Using Eq.~\eqref{eq: delTheta}, we obtain
\beq
&&\partial_{k_j}\left(d_i^{\Theta^\mu}\right)_{n}-\partial_{k_i}\left(d_j^{\Theta^\mu}\right)_{n}\nn\\
&=&\Re \sum_{n_1(\neq n)}\left[\partial_{k_j}(\mathcal{A}_i^k)_{nn_1}-\partial_{k_i}(\mathcal{A}_j^k)_{nn_1}\right]\Theta^\mu_{n_1n}\nn\\
&-&\Im \sum_{n_1(\neq n),n_2}\Theta^\mu_{n_2n}\left[(\mathcal{A}_i^k)_{nn_1}(\mathcal{A}_j^k)_{n_1n_2}-(\mathcal{A}_j^k)_{nn_1}(\mathcal{A}_i^k)_{n_1n_2}\right]\nn\\
&+&\Im \sum_{n_1(\neq n),n_2}\Theta^\mu_{n_1n_2}\left[(\mathcal{A}_i^k)_{nn_1}(\mathcal{A}_j^k)_{n_1n_2}-(\mathcal{A}_j^k)_{nn_1}(\mathcal{A}_i^k)_{n_1n_2}\right].\nn\\
\eeq
Using this, we express the last two terms in Eq~\eqref{eq: positionalshift_k2} as
\begin{widetext}
\beq
\label{eq: positionalshift_k3}
-\partial_{k_j}\left(d_i^{\Theta^\mu}\right)_{n}+2\Im \sum_{n_1(\neq n), n_2(\neq n)}(\mathcal{A}_i^k)_{nn_1}\Theta^\mu_{n_1n_2}(\mathcal{A}_j^k)_{n_2n}&=&-\partial_{k_i}\left(d_j^{\Theta^\mu}\right)_{n}-\Im \sum_{n_1(\neq n)}\Theta^\mu_{n_1n}\left[i\partial_{k_j}+(\mathcal{A}_j^k)_{n}-(\mathcal{A}_j^k)_{n_1}\right](\mathcal{A}_i^k)_{nn_1}\nn\\
&+&\Im \sum_{n_1(\neq n)}\Theta^\mu_{n_1n}\left[i\partial_{k_i}+(\mathcal{A}_i^k)_{n}-(\mathcal{A}_i^k)_{n_1}\right](\mathcal{A}_j^k)_{nn_1}\nn\\
&-&\Im \sum_{n_1(\neq n)}\Theta^\mu_{n}\left[(\mathcal{A}_j^k)_{nn_1}(\mathcal{A}_i^k)_{n_1n}-(\mathcal{A}_i^k)_{nn_1}(\mathcal{A}_j^k)_{n_1n}\right]\nn\\
&-&\Im \sum_{n_1(\neq n),n_2(\neq n,n_1)}\Theta^\mu_{n_2n}\left[(\mathcal{A}_j^k)_{nn_1}(\mathcal{A}_i^k)_{n_1n_2}
-(\mathcal{A}_i^k)_{nn_1}(\mathcal{A}_j^k)_{n_1n_2}\right],\nn\\
\eeq
\end{widetext}
with $\Theta^\mu_{n} \equiv \Theta^\mu_{nn}$. 
Using the identity
\beq
&&\partial_{k_j}(\mathcal{A}_i^k)_{nn_1}-\partial_{k_i}(\mathcal{A}_j^k)_{nn_1}\nn\\
&=&i\sum_{n_2}\left[(\mathcal{A}_j^k)_{nn_2}(\mathcal{A}_i^k)_{n_2n_1}-(\mathcal{A}_i^k)_{nn_2}(\mathcal{A}_j^k)_{n_2n_1}\right],
\eeq
we rewrite Eq.~\eqref{eq: positionalshift_k3} as
\beq
&-&\partial_{k_j}\left(d_i^{\Theta^\mu}\right)_{n}+2\Im \sum_{n_1(\neq n), n_2(\neq n)}(\mathcal{A}_i^k)_{nn_1}\Theta^\mu_{n_1n_2}(\mathcal{A}_j^k)_{n_2n}\nn\\
&=&-\left(\mathcal{B}_{ij}^{kk}\right)_n \Theta^\mu_{n}-\partial_{k_i}\left(d_j^{\Theta^\mu}\right)_{n}.
\eeq
Substituting this into Eq.~\eqref{eq: positionalshift_k2}, we finally obtain~\cite{xiao_conSC}
\beq
\label{eq: positionalshift_k4}
\frac{\partial (\delta \mathcal{A}_i^k)_{n}}{\partial (\partial_{r_{cj}}h_\mu)}&=&\left(\mathcal{B}_{ij}^{kk\Theta^\mu}\right)_n-\left(\mathcal{B}_{ij}^{kk}\right)_n \Theta^\mu_{n}-\partial_{k_i}\left(d_j^{\Theta^\mu}\right)_{n}.
\eeq

\section{Field variation and local density}
We now derive the expression for the local density of $\hat{\theta}^\mu$ using the effective action of the wave packet~\cite{xiao_berrycur}.
The effective action is a functional for the auxiliary field $\bm h$ and its field variation is defined as
\beq
\label{eq: fieldvariation}
\frac{\delta \mathcal{S}[\bm h(r_c)]}{\delta \bm h(r)}
=\lim_{\bm \epsilon \to 0} \frac{\mathcal{S}[\bm h(r_c)+{\bm \epsilon}\delta(r-r_c)]-\mathcal{S}[\bm h(r_c)]}{\bm \epsilon}.
\eeq
with $r_c\equiv (\bm r_c,t_c)$ and $\delta(r-r_c)\equiv \delta(\bm r-\bm r_c)\delta(t-t_c)$.
Because the wave function of the wave packet satisfies the Schr\"{o}dinger equation, we can express this variation as
\beq
\label{eq: fieldvariation1}
\frac{\delta \mathcal{S}[\bm h(r_c)]}{\delta \bm h(r)}&=&-\int dt_c \bigg\langle \Psi_{n\bm k}\bigg|\frac{\delta \hat{\mathcal{H}}(r_c)}{\delta \bm h(r)}\bigg|\Psi_{n\bm k}\bigg\rangle\nn\\
&=&-\big\langle \Psi_{n\bm k}\big|{\hat {\bm \theta}}\delta(\bm r-\bm r_c)\big|\Psi_{n\bm k}\big\rangle.
\eeq
In the second line of Eq.~\eqref{eq: fieldvariation1}, we used
\beq
\frac{\delta \hat{\mathcal{H}}(r_c)}{\delta \bm h(r)}={\hat {\bm \theta}}\delta(r-r_c)\,.
\eeq
Therefore, the local density of $\hat{\theta}^\mu$ under the $\bm h$-field is given by~\cite{dong_semiclassical}
\beq
\label{eq: thetaloc}
\theta^\mu_{\rm loc}(\bm r,t,\bm h)&=&\sum_n\int [d\bm k]d\bm r_c D_nf_n \big\langle \Psi_{n\bm k}\big|{\hat {\bm \theta}}\delta(\bm r-\bm r_c)\big|\Psi_{n\bm k}\big\rangle\nn\\
&=&-\sum_n\int [d\bm k]d\bm r_c D_nf_n \frac{\delta \mathcal{S}[\bm h(r_c)]}{\delta \bm h(r)},
\eeq
where $f_n(\bm k,\bm r_c)$ is the quasiparticle distribution function and $D_n$ is the density of states (DOS) in the phase space~\cite{xiao_DOS}. 
To evaluate this, we perform the variation of the Lagrangian
\beq
\mathcal{L}(h_\mu+\delta h_\mu)-\mathcal{L}(h_\mu)&=&\frac{\partial \mathcal{L}}{\partial h_\mu}\delta h_\mu+\frac{\partial \mathcal{L}}{\partial (\partial_{r_{ci}}h_\mu)}\left(\partial_{r_{ci}}\delta h_\mu \right)\nn\\
&+&\frac{\partial \mathcal{L}}{\partial (\partial_{t_{c}}h_\mu)}\left(\partial_{t_{c}}\delta h_\mu \right)\nn\\
&+&\frac{\partial \mathcal{L}}{\partial (\partial_{r_{ci}}\partial_{r_{cj}}h_\mu)}\left(\partial_{r_{ci}}\partial_{r_{cj}}\delta h_\mu \right),
\eeq
with $\delta h_\mu=\epsilon_\mu \delta(r- r_c)$.
Then, the local density of $\hat{\theta}^\mu$ becomes
\beq
\label{eq: thetaloc1}
\theta^\mu_{\rm loc}(\bm r,t,\bm h)&=&-\sum_n\int [d\bm k] D_nf_n \frac{\partial \mathcal{L}}{\partial h_\mu}\nn\\
&+&\partial_{r_i} \sum_n\int [d\bm k] D_nf_n \frac{\partial \mathcal{L}}{\partial (\partial_{r_{i}}h_\mu)}\nn\\
&+&\sum_n\int [d\bm k] D_nf_n \left[\frac{d}{dt}\frac{\partial \mathcal{L}}{\partial (\partial_t h_\mu)}\right]\nn\\
&-&\partial_{r_i} \sum_n\int [d\bm k] D_nf_n \frac{\partial \mathcal{L}}{\partial (\partial_t h_\mu)}\dot{r}_i\nn\\
&-&\partial_{r_i}\partial_{r_j} \sum_n\int [d\bm k] D_nf_n \frac{\partial \mathcal{L}}{\partial (\partial_{r_i}\partial_{r_j} h_\mu)}.
\eeq
Substituting Eq.~\eqref{eq: lagrangian}, we obtain
\beq
\label{eq: thetaloc2}
\theta^\mu_{\rm loc}(\bm r,t,\bm h)&=&\sum_n\int [d\bm k]D_nf_n\bigg[\partial_{h_\mu}\tilde{\epsilon}_{n\bm k}-\left(\tilde{\mathcal{B}}_{\mu i}^{hk}\right)_n \dot{k}_i\nn\\
&-&\left(\tilde{\mathcal{B}}_{\mu i}^{hr}\right)_n \dot{r}_i-\left(\tilde{\mathcal{E}}_{\mu}^{h}\right)_n \bigg]\nn\\
&-&\partial_{r_i}\sum_n\int [d\bm k]D_nf_n\left[(d_i^{\theta^\mu})_n-\dot{k}_j\frac{\partial (\delta \mathcal{A}^k_j)_n}{\partial(\partial_i h_\mu)}\right]\nn\\
&+&\partial_{r_i}\sum_n\int [d\bm k]D_nf_n\nn\\
&\times&\left[\left(\dot{r}_{l}\partial_{r_l}h_j+\partial_th_\nu\right)\frac{\partial (\delta \mathcal{A}^h_\nu)_n}{\partial(\partial_i h_\mu)}\right]\nn\\
&+&\partial_{r_i}\partial_{r_j}\sum_n\int [d\bm k]D_nf_n q_{ij}^{\theta^\mu}.
\eeq

In this paper, we consider the intrinsic (impurity-independent) current responses, and approximate the distribution function by the equilibrium Fermi distribution function $f_n(\bm k,\bm r)=f(\tilde{\epsilon}_{n\bm k})$.
Because the wave packet energy depends on the inhomogeneous $\bm h$-field, the distribution function and the state-resolved thermodynamic potential $g(\tilde{\epsilon}_{n\bm k})=T\log(1-f(\tilde{\epsilon}_{n\bm k}))$ are expanded as
\beq
f(\tilde{\epsilon}_{n\bm k})&=&f(\epsilon_{n\bm k})+\left(\partial_{\epsilon_{n\bm k}}f(\epsilon_{n\bm k})\right)\delta \epsilon_{n\bm k},\\
g(\tilde{\epsilon}_{n\bm k})&=&g(\epsilon_{n\bm k})+f(\epsilon_{n\bm k})\delta \epsilon_{n\bm k},
\eeq
with $\delta \tilde{\epsilon}_{n\bm k}=\tilde{\epsilon}_{n\bm k}-\epsilon_{n\bm k}$.
We now assume a steady state under a uniform temperature gradient $\theta^\mu_{\rm loc}(\bm r,t)=\theta^\mu_{\rm loc}(\bm r)$, and the gradient of the state-resolved thermodynamic potential is given by
\beq
\partial_{r_i}g(\tilde{\epsilon}_{n\bm k})\simeq \mathcal{S}(\tilde{\epsilon}_{n\bm k})\left(-\partial_{r_i}T\right)
\eeq
with the state-resolved entropy density $\mathcal{S}(\tilde{\epsilon}_{n\bm k})=-\partial g(\tilde{\epsilon}_{n\bm k})/\partial T$.
Using the Bianchi identity
\beq
\partial_{r_i}\left(\tilde{\mathcal{B}}_{\mu i}^{hk}\right)_n+\partial_{h_\mu}\left(\tilde{\mathcal{B}}_{\mu i}^{kr}\right)_n+\partial_{k_i}\left(\tilde{\mathcal{B}}_{\mu i}^{rh}\right)_n=0,
\eeq
we expand Eq.~\eqref{eq: thetaloc2} up to the second order in the spatial gradient 
\begin{widetext}
\beq
\label{eq: thetaloc3}
\theta^\mu_{\rm loc}(\bm r,\bm h)&=&\partial_{h_\mu} \sum_n\int [d\bm k]\left[1+\left(\tilde{\mathcal{B}}_{ii}^{kr}\right)_n\right]g(\tilde{\epsilon}_{n\bm k})\nn\\
&-&\partial_{r_i}\left[\sum_n\int [d\bm k]\left(\left(d_i^{\theta^\mu}\right)_n f(\epsilon_{n\bm k})+\left(\tilde{\mathcal{B}}_{i\mu}^{kh}\right)_n g(\tilde{\epsilon}_{n\bm k})-\frac{\partial (\delta \mathcal{A}_j^k)_n}{\partial (\partial_{r_i}h_\mu)}\mathcal{S}\left(\epsilon_{n\bm k}\right)\left(-\partial_{r_j}T\right)\right)\right]\nn\\
&+&\partial_{r_i}\partial_{r_j}\sum_n\int [d\bm k]\left[\left(q^{\theta^\mu}_{ij}\right)_nf\left(\epsilon_{n\bm k}\right)-\frac{\partial (\delta \mathcal{A}_j^k)_n}{\partial (\partial_{r_i}h_\mu)}g\left(\epsilon_{n\bm k}\right)\right]
-\sum_n\int [d\bm k]\left(\tilde{\mathcal{B}}_{\mu j}^{hk}\right)_n\mathcal{S}\left(\epsilon_{n\bm k}\right)\left(-\partial_{r_j}T\right)\nn\\
&-&\sum_n\int [d\bm k]\left[\left(\tilde{\mathcal{B}}_{\mu i}^{hk}\right)_n\left(\tilde{\mathcal{B}}_{ij}^{rk}\right)_n-\left(\tilde{\mathcal{B}}_{\mu i}^{hr}\right)_n\left(\tilde{\mathcal{B}}_{ij}^{kk}\right)_n-\left(\tilde{\mathcal{B}}_{\mu j}^{hk}\right)_n\left(\tilde{\mathcal{B}}_{ii}^{rk}\right)_n\right]\mathcal{S}\left(\epsilon_{n\bm k}\right)\left(-\partial_{r_j}T\right).
\eeq

We first consider the local equilibrium system.
In the absence of the temperature gradient, Eq.~\eqref{eq: thetaloc3} reduces to
\beq
\label{eq: thetaloc_leq}
\theta^\mu_{\rm loc,leq}(\bm r,\bm h)&=&\partial_{h_\mu} \sum_n\int [d\bm k]\left[1+\left(\tilde{\mathcal{B}}_{ii}^{kr}\right)_n\right]g(\tilde{\epsilon}_{n\bm k})-\partial_{r_i}\sum_n\int [d\bm k]\left[\left(d_i^{\theta^\mu}\right)_n f(\epsilon_{n\bm k})+\left(\tilde{\mathcal{B}}_{i\mu}^{kh}\right)_n g(\tilde{\epsilon}_{n\bm k})\right]\nn\\
&+&\partial_{r_i}\partial_{r_j}\sum_n\int [d\bm k]\bigg[\left(q^{\theta^\mu}_{ij}\right)_nf\left(\epsilon_{n\bm k}\right)-\frac{\partial (\delta \mathcal{A}_j^k)_n}{\partial (\partial_{r_i}h_\mu)}g\left(\epsilon_{n\bm k}\right)\bigg].
\eeq
Using
\beq
&&\left(\tilde{\mathcal{B}}_{ii}^{kr}\right)_n
\simeq \left(\tilde{\mathcal{B}}_{i\mu}^{kh}\right)_n(\partial_{r_i}h_\mu)-\frac{\partial (\delta \mathcal{A}_i^k)_n}{\partial (\partial_{r_j}h_\mu)}(\partial_{r_i}\partial_{r_j}h_\mu),
\eeq
we obtain
\beq
\label{eq: thetaloc_leq1}
\theta^\mu_{\rm loc,leq}(\bm r,\bm h)&\simeq& \sum_n\int [d\bm k]\partial_{h_\mu}\bigg[g(\tilde{\epsilon}_{n\bm k})+\left\{\left(d_i^{\theta^\mu}\right)_n f(\epsilon_{n\bm k})+\left(\tilde{\mathcal{B}}_{il}^{kh}\right)_n g(\tilde{\epsilon}_{n\bm k})\right\}(\partial_{r_i}h_\mu)\nn\\
&+&\left\{ \left(q^{\theta^\mu}_{ij}\right)_nf\left(\epsilon_{n\bm k}\right)-\frac{\partial (\delta \mathcal{A}_j^k)_n}{\partial (\partial_{r_i}h_\mu)}g(\epsilon_{n\bm k}) \right\}(\partial_{r_i}\partial_{r_j}h_\mu) \bigg]-\partial_{r_i}\sum_n\int [d\bm k]\left[\left(d_i^{\theta^\mu}\right)_n f(\epsilon_{n\bm k})+\left(\mathcal{B}_{i\mu}^{kh}\right)_n g(\tilde{\epsilon}_{n\bm k})\right]\nn\\
&+&\partial_{r_i}\partial_{r_j} \sum_n\int [d\bm k]\bigg[ \left(q^{\theta^\mu}_{ij}\right)_nf\left(\epsilon_{n\bm k}\right)-\frac{\partial (\delta \mathcal{A}_j^k)_n}{\partial (\partial_{r_i}h_\mu)}g(\epsilon_{n\bm k}) \bigg].
\eeq
At this stage, we turn off the $\bm h$-field in Eq.~\eqref{eq: thetaloc_leq1}, yielding
\beq
\label{eq: thetaloc_leq2}
\theta^\mu_{\rm loc,leq}(\bm r)&\equiv&\lim_{\bm h\to 0}\theta^\mu_{\rm loc,leq}(\bm r,\bm h)\nn\\
&=&\sum_n\int [d\bm k]\left(\theta^\mu\right)_nf(\epsilon_{n\bm k})
-\partial_{r_i}\sum_n\int [d\bm k]\left[\left(d_i^{\theta^\mu}\right)_n f(\epsilon_{n\bm k})+\left(\mathcal{B}_{i\mu}^{kh}\right)_n\bigg|_{\bm h\to 0} g(\epsilon_{n\bm k})\right]\nn\\
&+&\partial_{r_i}\partial_{r_j} \sum_n\int [d\bm k]\bigg[ \left(q^{\theta^\mu}_{ij}\right)_nf\left(\epsilon_{n\bm k}\right)-\frac{\partial (\delta \mathcal{A}_j^k)_n}{\partial (\partial_{r_i}h_\mu)}\bigg|_{\bm h\to 0}g(\epsilon_{n\bm k}) \bigg].
\eeq
Similarly, the local density of $\hat{\theta}^\mu$ in the nonequilibrium steady state can be calculated.
The result is given by~\cite{xiao_conSC}
\beq
\label{eq: thetaloc_neq}
\theta^\mu_{\rm loc}(\bm r)&\equiv&\lim_{\bm h\to 0}\theta^\mu_{\rm loc}(\bm r,\bm h)\nn\\
&=&\sum_n\int [d\bm k]\left(\theta^\mu\right)_nf(\epsilon_{n\bm k})
-\sum_n\int [d\bm k] \left(\mathcal{B}_{j\mu}^{kh}\right)_n\bigg|_{\bm h\to 0} \mathcal{S}(\epsilon_{n\bm k}) (-\partial_{r_j}T)
\nn\\
&-&\partial_{r_i}\sum_n\int [d\bm k]\left[\left(d_i^{\theta^\mu}\right)_n f(\epsilon_{n\bm k})+\left(\mathcal{B}_{i\mu}^{kh}\right)_n\bigg|_{\bm h\to 0} g(\tilde{\epsilon}_{n\bm k})\right]+\partial_{r_i}\sum_n\int [d\bm k] \frac{\partial (\delta \mathcal{A}_j^k)_n}{\partial (\partial_{r_i}h_\mu)}\bigg|_{\bm h\to 0}\mathcal{S}(\epsilon_{n\bm k}) (-\partial_{r_j}T)\nn\\
&+&\partial_{r_i}\partial_{r_j} \sum_n\int [d\bm k]\bigg[ \left(q^{\theta^\mu}_{ij}\right)_nf\left(\epsilon_{n\bm k}\right)-\frac{\partial (\delta \mathcal{A}_j^k)_n}{\partial (\partial_{r_i}h_\mu)}\bigg|_{\bm h\to 0}g(\epsilon_{n\bm k}) \bigg].
\eeq
\end{widetext}

\section{Equilibrium conserved spin current}
\label{sec: equilibrium spin current}
As described in the main text, the conserved spin current has two parts: the conventional spin current and the spin torque dipole.
By substituting $\hat{\theta}^\mu=\hat{J}^{s^\mu}_i$ into Eq.~\eqref{eq: thetaloc_leq2}, we obtain the equilibrium conventional spin current 
\beq
J_{i, {\rm leq}}^{s^\mu} &\simeq& \int [d\bm k]\left(\hat{J}^{s^\mu}_i\right)_nf(\epsilon_{n\bm k})\nn\\
&-&\partial_{r_l}\int [d\bm k]\left[\left(d_l^{J_{i}^{s^\mu}}\right)_n f(\epsilon_{n\bm k})+\left(\mathcal{B}_{li}^{kks^\mu}\right)_n g(\tilde{\epsilon}_{n\bm k})\right],\nn\\
\eeq
where the terms involving the second-order spatial gradient are neglected.
Here, $\left(\mathcal{B}_{li}^{kks^\mu}\right)_n$ represents the spin Berry curvature [Eq.~\eqref{spin_Berrycur}] and $\left(d_l^{J_{i}^{s^\mu}}\right)_n$ is the dipole moment of the conventional spin current, given by
\beq
\left(d_l^{J_{i}^{s^\mu}}\right)_n&=&\Re \sum_{n_1(\neq n)}(\mathcal{A}_l^k)_{nn_1}(J_{i}^{s^\mu})_{n_1n}\nn\\
&=&\Im \sum_{n_1(\neq n)}\frac{(\mathcal{A}_l^k)_{nn_1}(J_{i}^{s^\mu})_{n_1n}}
{\epsilon_{n\bm k}-\epsilon_{n_1\bm k}}.
\eeq
Next, by substituting $\hat{\theta}^\mu=\hat{\tau}^{\mu}=(1/i)[\hat{s}^\mu,\hat{\mathcal{H}}]$ into Eq.~\eqref{eq: thetaloc_leq2}, we obtain the equilibrium spin torque density
\begin{widetext}
\beq
\tau^{\mu}_{{\rm leq}} &=& -\partial_{r_i} \bigg[\sum_n\int [d\bm k]\left\{\left(d_i^{\tau^\mu}\right)_n f(\epsilon_{n\bm k})+\left(\mathcal{B}_{i\mu}^{kh^\tau}\right)_n\bigg|_{\bm h^\tau\to 0} g(\epsilon_{n\bm k})\right\}\nn\\
&-&\partial_{r_j}\sum_n \int [d\bm k] \left\{\left(q^{\theta^\mu}_{ij}\right)_nf\left(\epsilon_{n\bm k}\right)-\frac{\partial (\delta \mathcal{A}_j^k)_n}{\partial (\partial_{r_i}h^{\tau}_\mu)}\bigg|_{\bm h^\tau \to 0}g(\epsilon_{n\bm k})\right\} \bigg].
\eeq
Here, $\bm h^\tau$ denotes a conjugate field of the spin torque operator.
Because the local spin torque dipole density satisfies $\tau_{\rm loc}^{\mu}(\bm x) = -{\bm \nabla} \cdot \bm J_{\rm loc}^{\tau^\mu}$, its equilibrium value is given by
\beq
J^{\tau^\mu}_{i,{\rm leq}} &=& \sum_n\int [d\bm k]\left\{\left(d_i^{\tau^\mu}\right)_n f(\epsilon_{n\bm k})+\left(\mathcal{B}_{i\mu}^{kh^\tau}\right)_n\bigg|_{\bm h^\tau \to 0} g(\epsilon_{n\bm k})\right\}\nn\\
&-&\partial_{r_j} \sum_n\int [d\bm k] \left\{\left(q^{\tau^\mu}_{ij}\right)_nf\left(\epsilon_{n\bm k}\right)-\frac{\partial (\delta \mathcal{A}_j^k)_n}{\partial (\partial_{r_i}h^\tau_\mu)}\bigg|_{\bm h^\tau \to 0}g(\epsilon_{n\bm k})\right\} .
\eeq
\end{widetext}
From Eq.~\eqref{eq: dthetan}, the dipole moment of the spin torque is evaluated as
\beq
\left(d_i^{\tau^\mu}\right)_n&=&\Re \sum_{n_1(\neq n)}(\mathcal{A}_i^k)_{nn_1}i(\epsilon_{n_1\bm k}-\epsilon_{n\bm k})s^\mu_{n_1n}\nn\\
&=&-\Re \sum_{n_1}v^i_{nn_1}s^\mu_{n_1n}+v^i_ns_n^\mu\nn\\
&=&-\left(\hat{J}^{s^\mu}_i\right)_n+v^i_ns_n^\mu.
\eeq
Using
\beq
\left(\mathcal{A}_\mu^{h_\tau}\right)_{n_1n}&=&i\langle u_{n_1\bm k}|\partial_{h^\tau_\mu}u_{n\bm k}\rangle\nn\\
&=&\frac{i\langle u_{n_1\bm k}|\hat{\tau}^\mu|u_{n\bm k}\rangle}{\epsilon_{n\bm k}-\epsilon_{n_1\bm k}}\nn\\
&=&s_{n_1n}^\mu,
\eeq
we simplify $\left(\mathcal{B}_{i\mu}^{kh_\tau}\right)_n$ as
\beq
\left(\mathcal{B}_{i\mu}^{kh_\tau}\right)_n=\partial_{k_i}s_n^\mu.
\eeq
From Eqs.~\eqref{eq: quadrupole5} and~\eqref{eq: positionalshift_k4}, we obtain
\beq
\left(q_{ij}^{\tau^\mu}\right)_n
&=&-\frac{1}{2}\left(\mathcal{B}_{ij}^{kks^\mu}\right)_n+\left(d_i^{s^\mu}\right)_{n}v_n^j-\left(m_{ij}^{s^\mu}\right)_n,\\
\label{eq: positionalshift_st}
\frac{\partial (\delta \mathcal{A}_i^k)_{n}}{\partial (\partial_{r_{j}}h^\tau_\mu)}\bigg|_{\bm h^\tau \to 0}&=&\left(\mathcal{B}_{ij}^{kks^\mu}\right)_n-\left(\mathcal{B}_{ij}^{kk}\right)_n s^\mu_{n}-\partial_{k_i}\left(d_j^{s^\mu}\right)_{n}.
\eeq
Using these relations, we find that the equilibrium conserved spin current takes a circulating form~\cite{xiao_conSC,koyama}
\beq
\label{eq: localSC}
\bm{\mathcal{J}}_{\rm leq}^{s^\mu} &=& {\bm J}_{\rm leq}^{s^\mu}+{\bm J}_{\rm leq}^{\tau^\mu}\nn\\
&=&{\bm \nabla} \times {\bm M}^{s^\mu},
\eeq
with
\beq
M_l^{s^\mu}=\epsilon_{ijl}\sum_n\int [d\bm k]\left[\left(m_{ij}^{s^\mu}\right)_nf(\epsilon_{n\bm k})+\left(\mathcal{B}_{ij}^{kk}\right)_n s^\mu_{n}g(\epsilon_{n\bm k})\right].\nn\\
\eeq

\section{Derivation of $\alpha_{ij}^{s^\mu}$}
\label{sec: derivation SNC}
To accurately describe the current response to temperature gradients, it is necessary to eliminate the contribution of the circulating currents from the local current because the circulating currents can not be measured by transport experiments.
Hence, we define the transport (conserved) spin current as~\cite{smrcka1977transport,cooper_thermoelectric,xiao_ANE,qin_ATHE,shitade_torsion,sumiyoshi_ATHE,fujimoto_SNE}
\beq
\bm{\mathcal{J}}^{s^\mu}=\bm{\mathcal{J}}_{\rm loc}^{s^\mu}-{\bm \nabla} \times {\bm M}^{s^\mu}.
\eeq
From Eqs.~\eqref{eq: thetaloc_neq} and \eqref{eq: localSC}, the SNC is expressed as
\beq
\alpha_{ij}^{s^\mu}=\sum_n\int [d\bm k] \left[\left(\mathcal{B}_{ij}^{kks^\mu}\right)_n-\frac{\partial (\delta \mathcal{A}_i^k)_{n}}{\partial (\partial_{r_{j}}h^\tau_\mu)}\bigg|_{\bm h^\tau \to 0}\right]\mathcal{S}(\epsilon_{n\bm k}).\nn\\
\eeq
Using Eq.~\eqref{eq: positionalshift_st}, we finally obtain~\cite{xiao_conSC}
\beq
\alpha_{ij}^{s^\mu} = \sum_n \int [d\bm k]  \left[s_n^\mu \left(\mathcal{B}_{ij}^{kk}\right)_n+\partial_{k_j}\left(d_{i}^{s^\mu}\right)_n\right] \mathcal{S}(\epsilon_{n\bm k}).\nn\\
\eeq

\section{Equilibrium quasiparticle charge current}
The concept of the conserved current was recently applied to the charge current in superconductors.
In superconductors, the charge current $\hat{\bm J}=\left\{ \hat{\rho},\hat{v}(\bm k) \right\}$ is represented by the anticommutator of the charge operator $\hat{\rho}=\hat{\sigma}_z$ and the velocity operator $\hat{v}(\bm k)$.
Its time evolution obeys $d\rho/dt=-\bm \nabla \cdot \bm J^\tau$ and $\bm J^\tau$ accounts for a condensate backflow that ensures the conservation of the charge current~\cite{xiao_conSC}.

Similarly to Sec.~\ref{sec: equilibrium spin current}, by substituting $\hat{\theta}^\mu=\hat{\rho}, d\hat{\rho}/dt$ into Eq.~\eqref{eq: thetaloc_leq2}, we obtain the equilibrium quasiparticle charge current with a circulating form
\beq
\label{eq: localEC}
\bm{\mathcal{J}}_{\rm leq} &\equiv & {\bm J}_{\rm leq}+{\bm J}_{\rm leq}^{\tau}\nn\\
&=&{\bm \nabla} \times {\bm M},
\eeq
where the orbital magnetization is given by
\beq
M_l=\epsilon_{ijl}\sum_n\int [d\bm k]\left[\left(m_{ij}\right)_nf(\epsilon_{n\bm k})+\left(\mathcal{B}_{ij}^{kk}\right)_n \rho_{n}g(\epsilon_{n\bm k})\right].\nn\\
\eeq
Here, the quasiparticle charge is expressed as
\beq
\langle u_{m\bm p}|\hat{\rho}| u_{n\bm p}\rangle=
\begin{cases}
\rho_{n} & n=m,\\
(\rho)_{mn} & n\neq m.
\end{cases}
\eeq
$\left(m_{ij}\right)_n$ represents the orbital magnetic moment of quasiparticles, defined as
\beq
\left(m_{ij}\right)_n&=&\frac{1}{2}\Re \sum_{n_1(\neq n)}
\bigg[  
(\mathcal{A}_i^k)_{nn_1}\left(J_j\right)_{n_1n}-(\mathcal{A}_j^k)_{nn_1}\left(J_i\right)_{n_1n}
\bigg]\nn\\
&+&\frac{1}{2}\left[
\left(d^{\rho}_i\right)_{n}v_n^j-\left(d^{\rho}_j\right)_{n}v_n^i
\right].
\eeq
The quantity $\left(d^{\rho}_i\right)_{n}$ represents the charge dipole moment of quasiparticles, defined by
\beq
\left(d^{\rho}_i\right)_{n}&=&\Re \sum_{m(\neq n)} (\mathcal{A}_i^k)_{nm}(\rho)_{mn}.
\eeq
A nonzero dipole moment indicates that the charge center of the quasiparticle wave packet does not coincide with its probability center. 

\section{Derivation of $\alpha_{ij}$}
Equation \eqref{eq: localEC} takes a circulating form, allowing the definition of the transport charge current by subtracting the magnetization charge current from the local charge current induced by temperature gradients~\cite{smrcka1977transport,cooper_thermoelectric,xiao_ANE,qin_ATHE,shitade_torsion,sumiyoshi_ATHE,fujimoto_SNE}
\beq
\bm{\mathcal{J}}=\bm{\mathcal{J}}_{\rm loc}-{\bm \nabla} \times {\bm M}.
\eeq
In a similar manner to Sec.~\ref{sec: derivation SNC}, by substituting $\hat{\theta}^\mu=\hat{\rho}, d\hat{\rho}/dt$ into Eq.~\eqref{eq: thetaloc_neq}, we obtain
\beq
\alpha_{ij} = \sum_n \int [d\bm k]  \left[\rho_n \left(\mathcal{B}_{ij}^{kk}\right)_n+\partial_{k_j}\left(d^{\rho}_{i}\right)_n\right] \mathcal{S}(\epsilon_{n\bm k}).
\eeq

\bibliographystyle{apsrev4-1_PRX_style}

\bibliography{SNE.bib}
\end{document}